\documentclass[12pt]{article}

\usepackage{amsmath,dsfont,amssymb,graphicx} %use drftcite in draftmode
\usepackage{epsf}
\usepackage{pstricks}
\usepackage{cite}

\newcommand{\beq}{\begin{equation}}% can be used as {equation} or  {eqnarray}
\newcommand{\eeq}{\end{equation}}
\newcommand{\bes}{\begin{equation} \begin{split}}
\newcommand{\ees}{\end{split} \end{equation}}
\newcommand{\drawsquare}[2]{\hbox{%
\rule{#2pt}{#1pt}\hskip-#2pt%  left vertical
\rule{#1pt}{#2pt}\hskip-#1pt%  lower horizontal
\rule[#1pt]{#1pt}{#2pt}}\rule[#1pt]{#2pt}{#2pt}\hskip-#2pt%  upper horizontal
\rule{#2pt}{#1pt}}% right vertical
\newcommand{\Yfund}{\drawsquare{7}{0.6}}%  fundamental
\newcommand{\Ysymm}{\drawsquare{7}{0.6}\hskip-0.6pt%
   \drawsquare{7}{0.6}}%  symmetric second rank tensor
\DeclareMathOperator*{\Tr}{Tr}
\newcommand{\noSUSY}{\begin{picture}(25,0)(0,0)
        \put(0,0){\scriptsize $SUSY$}
        \put(0,0){\line(4,1){22}}
        \end{picture}}

\def\beqa{\begin{eqnarray}}
\def\eeqa#1{\label{#1}\end{eqnarray}}
\def\eeqan{\end{eqnarray}}
\def\CR{\nonumber \\ }
\def\leqn#1{(\ref{#1})}

\def\stacksymbols #1#2#3#4{\def\theguybelow{#2}
    \def\vp{\lower#3pt}
    \def\sp{\baselineskip0pt\lineskip#4pt}
    \mathrel{\mathpalette\intermediary#1}}

\def\intermediary#1#2{\vp\vbox{\sp
     \everycr={}\tabskip0pt
     \halign{$\mathsurround0pt#1\hfil##\hfil$\crcr#2\crcr
              \theguybelow\crcr}}}

\def\gsim{\stacksymbols{>}{\sim}{2.5}{.2}}
\def\lsim{\stacksymbols{<}{\sim}{2.5}{.2}}

\newcommand{\centeron}[2]{{\setbox0=\hbox{#1}\setbox1=\hbox{#2}\ifdim

\wd1>\wd0\kern.5\wd1\kern-.5\wd0\fi \copy0

\kern-.5\wd0\kern-.5\wd1\copy1\ifdim\wd0>\wd1
                                       \kern.5\wd0\kern-.5\wd1\fi}}
\newcommand{\ltap}{\>\centeron{\raise.35ex\hbox{$<$}}
                               {\lower.65ex\hbox{$\sim$}}\>}
\newcommand{\gtap}{\>\centeron{\raise.35ex\hbox{$>$}}
                               {\lower.65ex\hbox{$\sim$}}\>}

\newcommand\ZZ{\hbox{\zfont Z\kern-.4emZ}}
\font\zfont = cmss10 %scaled \magstep1

\def\gp3{g^\prime_3}

\begin{document}

%%%%%%%%%%%%%%%%%%%%%%%%%%%%%%%%%%%%%%%%%%%%%%%%%%%%%%
\begin{titlepage}

\vskip.5cm

%%%%%%%%%%%%%%%%%%%%%%%%%%%
\begin{center}
{\huge \bf A Weakly Coupled Ultraviolet Completion of the Littlest Higgs with T-parity}
\vskip.1cm
\end{center}
\vskip0.2cm

%%%%%%%%%%%%%%%%%%%%%%%%%%%
\begin{center}
{\bf Csaba Cs\'aki$^{a,b}$, Johannes Heinonen$^{a,b}$, Maxim Perelstein$^a$ \\  and Christian Spethmann$^a$}
\end{center}

\vskip 8pt

%%%%%%%%%%%%%%%%%%%%%%%%%%%
\begin{center}
{\it $^a$ Institute for High Energy Phenomenology\\
Newman Laboratory of Elementary Particle Physics\\
Cornell University, Ithaca, NY 14853, USA } \\
\vspace*{0.3cm}
{\it $^b$ Kavli Institute for Theoretical Physics\\
University of California, Santa Barbara, CA 93106-4030, USA } \\
\vspace*{0.3cm}
{\tt  csaki,heinonen,maxim@lepp.cornell.edu; cs366@cornell.edu}
\end{center}

\vglue 0.3truecm

%%%%%%%%%%%%%%%%%%%%%%%%%%%
\begin{abstract}
\vskip 3pt \noindent
We construct a weakly coupled, renormalizable ultraviolet completion of the
Littlest Higgs model with T-parity (LHT), based on an $SU(5) \times SU(2) \times U(1)$
gauge theory with a discrete $Z_2$ symmetry. Our model reproduces the complete
structure of the LHT below the 10 TeV scale, including the collective
symmetry breaking mechanism which solves the little hierarchy problem.
The model is manifestly free of anomalies, including both gauge/gravitational
anomalies and anomalies involving T-parity. At the TeV scale, the model
contains additional states not present in the LHT. We estimate the impact of
these states on precision electroweak observables, and show that the model
is realistic. We also discuss how our model can be embedded into a
supersymmetric theory or a five-dimensional setup with a warped extra
dimension, stabilyzing the hierarchy between the 10 TeV and the Planck
scale.

\end{abstract}

\end{titlepage}

%%%%%%%%%%%%%%%%%%%%%%%%%%%%%%%%%%%%%%%%%%%%%%%%%%%%%%
%%%%%%%%%%%%%%%%%%%%%%%%%%%%%%%%%%%%%%%%%%%%%%%%%%%%%%
\section{Introduction}
\label{sec:Intro} \setcounter{equation}{0} %\setcounter{footnote}{0}
%%%%%%%%%%%%%%%%%%%%%%%%%%%%%%%%%%%%%%%%%%%%%%%%%%%%%%
%%%%%%%%%%%%%%%%%%%%%%%%%%%%%%%%%%%%%%%%%%%%%%%%%%%%%%

One of the most pressing issues facing particle theory is the little
hierarchy problem. On the one hand, electroweak precision
measurements at LEP and the Tevatron seem to indicate the existence
of a weakly coupled light (below 200 GeV) Higgs boson. This Higgs
would be unstable against large radiative corrections, and one would
expect new physics at or below the TeV scale to stabilize the Higgs
potential. On the other hand, the same electroweak precision
measurements have failed to provide any indirect evidence for such
physics. For the case of supersymmetry (SUSY), a natural minimal model
should have already been discovered at LEP2 or the Tevatron:
 null results of superpartner and Higgs
searches imply that a  fine-tuning of order 1\% or worse is required
to accommodate the data, which is the particular incarnation of the
little hierarchy problem for SUSY.

The motivation for Little Higgs (LH) models is to solve this issue by
pushing the scale of new physics that solves the ``large'' (weak/Planck) 
hierarchy
problem up to 10 TeV, and provide a rationale for the cancelation of
the remaining quadratic divergences in the Higgs mass between 1 TeV and 10 
TeV. This is achieved by interpreting the Higgs as an
approximate Goldstone boson corresponding to a spontaneously broken
global symmetry of the electroweak sector. Gauge and Yukawa
couplings of the Higgs must break the global symmetry explicitly;
however, if this breaking is ``collective'' (meaning that no single
coupling breaks all of the symmetry responsible for keeping the
Higgs light), the extended theory can remain perturbative until the
10 TeV scale without fine-tuning~\cite{collective}. Several explicit
realizations of this idea have appeared in the
literature~\cite{lhreviews}. Models with T-parity are especially promising,
since they can be consistent with precision electroweak constraints without
need for fine tuning in the Higgs mass~\cite{Tparity}.
In this paper, we will focus on the
Littlest Higgs model with T-parity (LHT)~\cite{LHT}, which is a
fully realistic example of this class.

Like all existing Little Higgs models, the LHT has been constructed
as an effective field theory, valid below the cutoff scale of order
10 TeV. This is sufficient to discuss the model's consistency with
precision electroweak data~\cite{PEW,As}, its signatures at the
Tevatron~\cite{tev} and the LHC~\cite{JP,bel}, and the dark matter
candidate that naturally emerges in this model~\cite{JP,DM,As}.
However, in order to really complete the program outlined above one
needs to find the ultraviolet (UV) completion of these models, {i.e.} embed it into a more fundamental theory valid at higher
scales, possibly all the way up to the scale of grand unification
(GUT) or the Planck scale. The main aim of this paper is to present
such a construction. As with most BSM models, there are two
possibilities. The UV completion may be a strongly coupled theory,
which happens to produce the LHT as its effective theory below the
confinement scale of 10 TeV, or the UV completion remains
perturbative, and the LHT emerges as a low-energy description of a
renormalizable weakly coupled gauge theory. Here we choose to follow
the second possibility, that is we present a linear UV completion of
the LHT. In this approach, one needs to introduce supersymmetry to stabilize
the hierarchy between the 10 TeV scale and the GUT/Planck scale;
however, since SUSY is broken at 10 TeV, the model is free of the
fine-tuning plaguing the MSSM. Alternatively one can have a Kaluza-Klein (KK)
tower of a warped extra dimension starting at 10 TeV, which would also
stabilize the large hierarchy. Our model explains the appearance and
radiative stability of the global symmetry structure of the LHT,
which at first sight appears rather unnatural. Furthermore, the
model is manifestly free of anomalies, including both the familiar
gauge/gravitational anomalies and the anomalies involving T-parity.
Thus, the anomaly-induced T-parity violating operators, which
recently received some attention in the
literature~\cite{Hill2,Hill_talk}, are completely absent in our
model  and T-parity is an exact symmetry, at least as long as
gravitational effects can be ignored. This illustrates the point
that the existence of these operators depends crucially on the
nature of the ultraviolet completion of the LH model. This has also
been emphasized very recently in~\cite{KY}, where it was also
pointed out the UV completions with anomalous T-parity are unlikely
to have the correct vacuum alignment.

Before presenting our model, let us briefly  comment on its relation
to previous work in this area. UV completions of the Littlest Higgs
model have been until now based on either a strongly interacting
theory or equivalently a warped extra dimension at the 10 TeV scale.
Models without T-parity have been
constructed~\cite{LHUV,ThalerYavin}, while recently an attempt to incorporate
a discrete parity based on two throats of warped dimensions
was presented in~\cite{Agashe:2007jb}. Our model is based on
conventional, four-dimensional and perturbative physics, making it
much easier to incorporate T-parity and to analyze anomalies.
Supersymmetric ultraviolet completions of an alternative LH model,
the ``simplest'' little Higgs, have also appeared in the
literature~\cite{superlittle1,superlittle2}. However, in those models
the electroweak precision constraints are so strong that one has to
assume that SUSY is broken at the weak scale, and the LH scale is
much higher. The role of the Little Higgs mechanism is to solve the
little hierarchy problem within SUSY. In contrast, in our model the
LH partners appear first, and SUSY is irrelevant until the 10 TeV
scale. At the LHC, our model would look like the familiar LHT, with
a few extra states. We will also present an extra
dimensional model that is reminiscent of the structure of the
minimal composite Higgs (MCH) models of~\cite{ACP}, in which the Higgs will
appear as the zero mode of the $A_5$ bulk gauge fields, which will
pick up a finite radiatively generated potential. The main
difference between the model presented here and the MCH models is
that we will have the T-odd little Higgs partners appearing at the 1
TeV scale, which will allow us to push the KK mass scale of the
theory to 10 TeV without fine-tuning. Thus the KK tower only plays a role of UV
completing the theory above 10 TeV and stabilizing the hierarchy
between 10 TeV and the Planck scale, but it is not used to cut off
the 1-loop quadratic divergences between 1 and 10 TeV.

The paper is organized as follows. We first construct a four-dimensional,
non-super\-symmetric, renormalizable model which reduces to the LHT (plus
a few extra states) below the 10 TeV scale. We discuss the bosonic (gauge
and scalar) sector of the model in section~\ref{sec:bosons}, and show how to
incorporate fermions in section~\ref{sec:fermions}. In
section~\ref{sec:anomalycancel}, we extend the model to achieve complete
anomaly cancelation, including anomalies involving T-parity. In
section~\ref{sec:HEcompletions}, we discuss how the hierarchy between the
10 TeV scale and the Planck scale can be stabilized by either
supersymmetrizing the model or embedding it into a theory with a warped fifth
dimension \`a la Randall and Sundrum\cite{RS}. In section~\ref{sec:constraints}, we
estimate the precision electroweak constraints on the model, and show that
the model is realistic. In section~\ref{sec:LHMinlsm}, we show
by an explicit diagrammatic calculation how the little Higgs cancelations
occur in our renormalizable model. Finally, section~\ref{sec:conclusions}
contains our conclusions.

%%%%%%%%%%%%%%%%%%%%%%%%%%%%%%%%%%%%%%%%%%%%%%%%%%%%%%
%%%%%%%%%%%%%%%%%%%%%%%%%%%%%%%%%%%%%%%%%%%%%%%%%%%%%%
\section{The Scalar/Gauge Sector for $SU(5)\times SU(2) \times U(1)$}
\label{sec:bosons}  \setcounter{equation}{0} %\setcounter{footnote}{0}
%%%%%%%%%%%%%%%%%%%%%%%%%%%%%%%%%%%%%%%%%%%%%%%%%%%%%%
%%%%%%%%%%%%%%%%%%%%%%%%%%%%%%%%%%%%%%%%%%%%%%%%%%%%%%

The bosonic (scalar and gauge) degrees of freedom of the LHT model are
described by a gauged non-linear sigma model (nl$\sigma$m). The scalars are the
Goldstone
bosons of  the global symmetry breaking $SU(5)\to SO(5)$. The symmetry-breaking
vev (or condensate) is in the symmetric representation {\bf 15} of the $SU(5)$.
The symmetry breaking scale $f_S$ is assumed to be about 1 TeV. To incorporate
the gauge degrees of freedom, an $[SU(2)\times U(1)]^2$ subgroup of the
$SU(5)$ is gauged; for the fundamental representation, the gauged subgroup of
$SU(5)$ is spanned by the generators
\begin{align}
Q_1^a =  \left( \begin{array}{c|c|c} \tau^a & &  \\ \hline & 0  & \\ \hline & & 0  \end{array} \right),  \qquad
Y_1 = \frac{1}{10}  \left( \begin{array}{cc|c|cc} 3 & & & &  \\ & 3 & & & \\ \hline & & -2 &  & \\ \hline & & & -2 & \\  & & & & -2  \end{array} \right) \\
\text{and} \quad Q_2^a =  \left( \begin{array}{c|c|c} 0 & &  \\ \hline & 0  & \\ \hline & & - \tau^{aT}  \end{array} \right),  \qquad  Y_2 = \frac{1}{10}  \left( \begin{array}{cc|c|cc} 2 & & & &  \\ & 2 & & & \\ \hline & & 2 &  & \\ \hline & & & -3 & \\  & & & & -3  \end{array} \right)
\end{align}
where $\tau^a=\sigma^a/2$. Below $f_S$, the gauge symmetry is reduced to the
diagonal
$SU(2)\times U(1)$, which is identified with the Standard Model (SM)
electroweak gauge group $SU(2)_L \times  U(1)_Y$. Under this group, the physical (uneaten) Goldstones
decompose into a weak doublet, identified with the SM Higgs, and a weak
triplet. The Higgs mass is protected from a one-loop quadratic divergence by
the collective symmetry breaking mechanism. The nl$\sigma$m is an effective theory
valid up to the scale $\Lambda\sim 4\pi f_S \sim 10$ TeV. For a more detailed
description of the LHT model, see Refs.~\cite{LHT,PEW,JP}.

The first step to a weakly coupled UV completion of the LHT is to replace the
nl$\sigma$m with a {\it linear} sigma model with the same symmetry breaking
structure. This model contains a single scalar field $S$, transforming as
{\bf 15} of $SU(5)$, which is assumed to get a vev
\beq
\langle S \rangle \,=\, f_S \left( \begin{array}{ccc} & & \mathds{1} \\ & 1 & \\ \mathds{1} & & \end{array} \right)\,,
\label{svev1}
\eeq
where $f_S\sim 1$ TeV. The Lagrangian is simply
\beq
{\cal L}_{\rm lin} = \tfrac{1}{8} |D_\mu S|^2 \,-\, V(S)\,,
\label{lin}
\eeq
where $D_\mu$ is the covariant derivative, and
the renormalizable potential $V(S)$ is assumed to
lead to an $S$ vev of the form~\leqn{svev1}.
We will not need to specify this and other scalar potentials explicitly in
this paper, for an example of a possible
potential for $S$ see eq.~\eqref{potentialforS}. The excitations around the
vacuum~\leqn{svev1} can be parametrized as
\beq
S =  \langle S \rangle \
 +  i \left( \begin{array}{ccc}
        \phi_S                  &      \sqrt{2}\, h_S         &   \chi_S + \frac{\eta_S}{\sqrt{5}}\\
        \sqrt{2}\,h_S^T                   &  -\frac{4\eta_S}{\sqrt{5}}    &   \sqrt{2}\,h_S^\dagger         \\
        \chi_S^T + \frac{\eta_S}{\sqrt{5}}&     \sqrt{2}\,h^*_S       &   \phi_S^\dagger
    \end{array}  \right)
 + \left( \text{radial modes} \right),
\eeq
where $\chi_S$ is a hermitan, complex 2$\times$2 matrix, $\eta_S$ a real
singlet, $\phi_S$ a complex, symmetric 2$\times$2 matrix and $h_S$ a complex
doublet, which will be identified with the SM Higgs. These fields are
pseudo-Goldstone bosons  (they would be exact Goldstone bosons, if the gauge couplings were taken to zero). They contain 14 degrees of freedom, corresponding to the number of $SU(5)$ generators broken by the $S$ vev. The other 16 degrees of freedom in $S$, the ``radial'' modes, obtain masses $\sim c f_S$, where $c$ are
order-one numbers determined by the coupling constants in $V(S)$.
Integrating out the radial modes reproduces the nl$\sigma$m description of the LHT, independent of the details of $V(S)$. This is guaranteed by the Coleman-Wess-Zumino theorem~\cite{CWZ}. In particular, the crucial feature of the LHT nl$\sigma$m is the special structure of the Higgs coupling to gauge fields, which guarantees the absence of a quadratic divergence in the Higgs mass at one loop. In section~\ref{sec:LHMinlsm}, we show by an explicit calculation how this structure emerges from the linear sigma model.

The model defined by eq.~\leqn{lin} is of course renormalizable, and can be
valid up to an arbitrarily high scale, for example the Planck scale. In this
sense, it is a viable UV completion of (the bosonic sector of) the LHT.
However, it has two significant shortcomings:

\begin{itemize}

\item The symmetry structure of this model is {\it very
unnatural}. Because gauge interactions break the global $SU(5)$ explicitly,
renormalization-group evolution generates $SU(5)$-violating operators in the
Lagrangian. In the LHT model, the global $SU(5)$ has to be a good symmetry at
the 10 TeV scale. This would require the linear model to contain a very
special combination of $SU(5)$-violating terms at the Planck scale, finely
tuned just so that the $SU(5)$ is miraculously restored at 10 TeV.

\item SM fermions {\it cannot} be incorporated in this model in a way
consistent with T-parity. T-parity requires that for every field transforming
under one of the two $SU(2)\times U(1)$ gauge groups of the LHT model, there
must be another field transforming in the same way under the other
$SU(2)\times U(1)$. Since the SM weak group is the diagonal combination of
the two $SU(2)$ factors, this means that the model must have an even number of
weak doublets of the same hypercharge and color charge. Therefore this model
cannot lead to the chiral fermion content of the SM in the low energy limit.

\end{itemize}

To avoid the first problem, we would like to start at high energies with a
model in which the full $SU(5)$ is promoted to a {\it gauge} symmetry.
Further, to incorporate chirality, we must enlarge the gauge structure to
contain an odd number of gauged $SU(2)$ factors. The most obvious and
easiest choice is to add one extra gauge $SU(2)$. As we will see below,
obtaining the correct hypercharge
assignments for all SM fermions also requires an additional $U(1)$ gauge
group. 

Thus, the full gauge group of our model, at high energies, is
\beq
SU(5)\times SU(2)_3\times U(1)_3,
\eeq
where we labeled the extra
$SU(2)\times U(1)$ factor with a subscript ``3'' to distinguish it from the
$[ SU(2)\times U(1)]^2$ subgroup of the $SU(5)$ that survives below 10 TeV.
To break the $[ SU(2)\times U(1)]^3$ subgroup to the SM electroweak gauge 
group, we also need additional bifundamental scalars under $SU(5)\times 
SU(2)_3$, $K_1$ and $K_2$, which will acquire the appropriate vevs (see 
eq.~\eqref{bvevs}).

\begin{table}
\begin{center}
\begin{equation*}
\begin{array}{|c|ccc|} \hline
&  SU(5)  &  SU(2)_3   &  U(1)_3  \\ \hline
 \Phi_{1,2}  & Adj & 1 & 0\\
 S  &  \Ysymm  & 1 &0 \\
 K_1  &  \Yfund  &  \Yfund  & -1/2 \\
 K_2  &  \overline{\Yfund}  &  \Yfund  & -1/2 \\ \hline
\end{array}
\end{equation*}
\caption{Scalar fields and their gauge charge assignments.}
\label{tab:scalars}
\end{center}
\end{table}

To reproduce the symmetries of the LHT model at low energies, we introduce a
set of scalar fields, summarized in Table~\ref{tab:scalars}. At the 10 TeV
scale, the $\Phi$ fields get vevs of the form
\begin{equation}
\langle \Phi_1 \rangle = f_\Phi \left( \begin{array}{ccccc} -3 \\ & -3
\\ && 2 \\ &&&2 \\ &&&&2 \end{array} \right)\,,
~~~\langle \Phi_2 \rangle = f_\Phi \left( \begin{array}{ccccc} 2 \\ & 2
\\ && 2 \\ &&&-3 \\ &&&&-3 \end{array} \right)\,
\label{phivev}
\end{equation}
where $f_\Phi\sim 10$ TeV. These vevs break the $SU(5)$ down to
$[SU(2)\times U(1)]^2$, the gauge group of the LHT model, and leave
the $SU(2)_3\times U(1)_3$ unbroken. If the scalar potential has the
form 
\beq 
V = V(\Phi_1,\Phi_2) + V(S,K_1,K_2)\,, \label{Vfact} 
\eeq
so that there are no direct couplings between $\Phi$'s and other
scalars, the model will possess an $SU(5)$ global symmetry below 10
TeV, broken only by gauge interactions. This is the idea that was
first emplyed in the context of $SU(6)$ GUT models in~\cite{SU6}, and
also in the "simplest little Higgs" model in~\cite{simplest}. With this
assumption, the full gauge/global symmetry structure of the LHT is reproduced.
Of course, this construction is only natural, if there is a symmetry
reason for the absence of direct potential couplings between
$\Phi$'s and the other scalars. In section~\ref{sec:HEcompletions},
we will show that the $\Phi$-vevs can be stabilized at the 10 TeV
scale, either by supersymmetrizing the model or by embedding it into
a five-dimensional model with warped geometry. In both cases, the
couplings between $\Phi$ and the other scalars can be naturally
suppressed.

At the 1 TeV scale, the field $S$ gets a vev given in eq.~\leqn{svev1}, while
the bifundamental fields get vevs
\beq
\langle K_1 \rangle  = f_K \left(
\begin{array}{cc} 1 & \\ & 1 \\ &  \\ & \\ &
\end{array}
\right),~~~~
\langle K_2 \rangle  = f_K \left(
\begin{array}{cc} & \\ & \\ &  \\ 1 & \\ & 1
\end{array}
\right)\,,
\label{bvevs}
\eeq
where $f_K\sim 1$ TeV. Together, these vevs break the
$[SU(2)\times U(1)]^3$ gauge symmetry down to a single $SU(2)\times U(1)$,
identified with the SM. The unbroken generators are simply
$Q_D^a = Q_1^a+Q_2^a+Q_3^a$ and $Y_D = Y_1 + Y_2 + Y_3$.

The global symmetry breaking by the $K$-vevs results in
additional pseudo-Goldstone bosons. We will assume that the tree-level
scalar potential does not contain direct couplings between the fields:
$V=V(S)+V(K_1,K_2)$.
With this assumption, the
Goldstones contained in different fields do not mix. Most of the Goldstones
are not protected by the collective
symmetry breaking mechanism. They will therefore receive quadratically
divergent masses at the one-loop level from gauge loops, and their masses
are in the TeV range. The only exceptions are the SM Higgs $h_S$, and a set 
of three real Goldstones transforming as a real triplet under the SM $SU(2)$ 
gauge group. Two of these triplets are eaten by the heavy $SU(2)$
gauge bosons, while the third one remains physical. The physical mode is a 
linear combination of the Goldstones coming from $S$, $K_1$ and $K_2$.
\begin{figure}
 \centering
  \includegraphics[width=7cm,keepaspectratio=true]{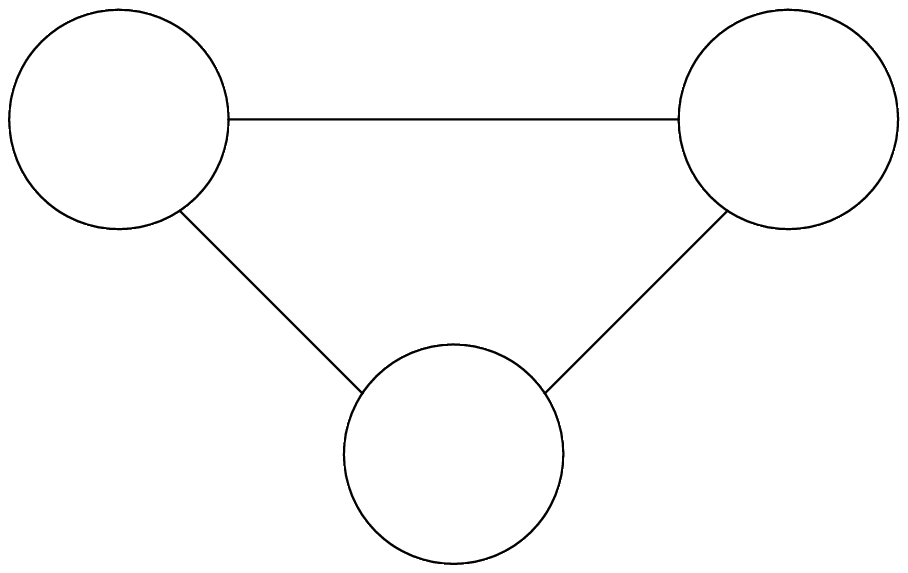}
  \put(-188,97){$SU(2)_1$}
  \put(-45,97){$SU(2)_2$}
  \put(-117,25){$SU(2)_3$}
  \put(-105,104){$S$}
  \put(-155,50){$K_1$}
  \put(-50,55){$K_2$}
  \caption{The gauge symmetries and scalar field content of the model below
the 10 TeV scale.}
\label{fig:moose}
\end{figure}
In fact, one can think of our model below 10 TeV as a three-site
deconstruction of a five-dimensional model, with the moose diagram shown in
Fig.~\ref{fig:moose}. In this picture, the light triplet mode is simply the
counterpart of $A_5$, and can
only receive a mass from non-local effects due to compactification. However,
the Yukawa couplings of our model (discussed in the following section) do not
have such an ``extra-dimensional'' structure, and the triplet mass is {\it not}
protected from the one-loop diagrams involving the Yukawas. Thus, this mode
will also receive a TeV-scale mass. The only pseudo-Goldstone protected by the
collective symmetry mechanism is the SM Higgs.

In addition to the gauge symmetries, we impose that the model is invariant
under a discrete T-parity, which acts on the gauge and scalar fields as
follows:
\beqa
W_{SU(5)} &\rightarrow& \Omega (W_{SU(5)}) \Omega^\dagger\,,\CR
W_{SU(2)} &\rightarrow& \omega (W_{SU(2)}) \omega^\dagger = W_{SU(2)}\,,\CR
B_{U(1)} &\rightarrow& B_{U(1)}\,,\CR
\Phi_1 &\leftrightarrow& \Omega \Phi_2 \Omega^\dagger\,, \CR
S  & \rightarrow&  \Omega  S^\dagger  \Omega^T \,, \CR
K_1  & \leftrightarrow&  \Omega  K_2 \omega^T\,,
\eeqa{Tparity}
where $W_{SU(5)}$, $W_{SU(2)}$ and
$B_{U(1)}$ are the SU(5), SU(2)$_3$ and U(1)$_3$ gauge fields, respectively, and
\beq
\Omega = \left( \begin{array}{ccc} & & -\mathds{1}Ê\\ & 1 & \\ -\mathds{1} & & \end{array} \right) \quad \text{ and } \quad \omega = -\mathds{1}.
\eeq
Note that $\Omega \in SU(5)$ and $\omega \in SU(2)$.
The kinetic terms are automatically invariant under this parity, while the
scalar potential must be restricted to the terms consistent with it.
The vevs in eqs.~\leqn{svev1},~\leqn{phivev} and~\leqn{bvevs} do not break
T-parity. It is easy to check that the T-parity defined in this way acts in
the desired way on the fields of the LHT model: the two $SU(2)\times U(1)$
factors inside the $SU(5)$ are interchanged, the Higgs boson $h_S$ is T-even,
while the weak triplet is T-odd, as required by precision electroweak fits.

Now, let us discuss the spectrum of the bosonic states. Sixteen out of the 24
$SU(5)$ gauge bosons get masses at the 10 TeV scale. These states are too heavy
to have any phenomenological consequences, and we will not discuss them
further. Below 10 TeV, we have three sets of $SU(2)$ gauge bosons:
\begin{align} \label{su2masses}
m_{W_{SM}}^2 & =  0: &  W_{SM} & =  \tfrac{1}{\sqrt{2 g_3^2+g_5^2}}
[ g_3(W_1+W_2)+g_5W_3]  \nonumber \\
m_{W_\text{even}}^2 & = \tfrac{g_5^2+2g_3^2}{4} f_K^2 :  & W_\text{even} & =
\tfrac{1}{\sqrt{2g_5^2+4g_3^2}}[g_5(W_1+W_2)-2g_3W_3] \\
m_{W_\text{odd}}^2 & =  \tfrac{g_5^2}{4}(2 f_S^2 +f_K^2): & W_\text{odd} &  =
\tfrac{1}{\sqrt{2}}[W_1-W_2]\,, \nonumber
\end{align}
as well as three $U(1)$ bosons:
\begin{align} \label{u1masses}
m_{B_{SM}}^2 & =  0: & B_{SM} & = \tfrac{1}{\sqrt{2 {g'_5}^2+{g'_3}^{2}}} [ g'_3(B_1+B_2)+g'_5 B_3] \nonumber \\
m_{B_\text{even}}^2 & =  \tfrac{{g'_5}^2+2{g'_3}^2}{4} f_K^2  :&  B_\text{even} & =\tfrac{1}{\sqrt{2{g'_5}^2+4{g'_3}^2}}[g'_5 (B_1+B_2)-2g'_3B_3] \\
m_{B_\text{odd}}^2 & = \tfrac{{g'_5}^2}{100}(10 f_S ^2 +f_K^2):& B_\text{odd} &  = \tfrac{1}{\sqrt{2}}[B_1-B_2]. \nonumber
\end{align}
Here $g_5$, $g_3$ and $g'_3$ are the $SU(5)$, $SU(2)_3$ and $U(1)_3$ coupling constants, respectively, and in proper normalization $g_5' = \sqrt{5/3} \, g_5$.

Note that the model contains a set of T-even gauge bosons at the TeV scale, due to the presence of an extra $SU(2)\times U(1)$ gauge factor, which is T-even. These states can be problematic for electroweak precision constraints, but are inevitable in our model. However, they do not participate in the cancelation of the quadratic divergences in the Higgs boson mass. Therefore, they can be substantially heavier than the T-odd states, without spoiling naturalness. This occurs if $g'_3, g_3\gg g_5$; if the T-odd states are at 1 TeV, requiring that $g'_3, g_3\sim$ 3-5 $g_5$ is sufficient to avoid precision electroweak constraints, and the model remains weakly coupled, but for these parameters, the Weinberg angle is fixed at a wrong value: $\sin^2\theta_W = 5/8$ in the limit $g'_3, g_3\gg g_5$. However, as we will discuss in section~\ref{sec:topsector}, reproducing the top sector of the LHT from a renormalizable model will require introduction of additional scalar vevs at the TeV scale, which will affect the gauge boson spectrum. It turns out that in the full model the correct value of the Weinberg angle can be easily reproduced without conflict with precision electroweak data, as we will show in detail in section~\ref{sec:constraints}.

%%%%%%%%%%%%%%%%%%%%%%%%%%%%%%%%%%%%%%%%%%%%%%%%%%%%%%
%%%%%%%%%%%%%%%%%%%%%%%%%%%%%%%%%%%%%%%%%%%%%%%%%%%%%%
\section{The Fermion Sector}
\label{sec:fermions}  \setcounter{equation}{0} %\setcounter{footnote}{0}
%%%%%%%%%%%%%%%%%%%%%%%%%%%%%%%%%%%%%%%%%%%%%%%%%%%%%%
%%%%%%%%%%%%%%%%%%%%%%%%%%%%%%%%%%%%%%%%%%%%%%%%%%%%%%
In this section we describe the fermion sector of our model that contains the SM fermions plus a number of heavier states. Our convention is to write all fermion fields as left-handed two-component  spinors.

%%%%%%%%%%%%%%%%%%%%%%%%%%%%
\subsection{The SM fermions}

It is straightforward to include the SM $SU(2)_L$ singlets as T-even  fermionic singlets, $u_R, d_R$ and $e_R$. (The SM generation index will be omitted throughout this paper.) For each SM doublet, we introduce two fermions in the representations {\bf 5} and $\overline{ \mathbf  5}$ of $SU(5)$
\beq \label{su5fermions}
\Psi_1 =
\left( \begin{array}{c} \psi_1 \\ U_{L1} \\ \chi_1 \end{array} \right)
\text{ and } \Psi_2 = \left( \begin{array}{c} \chi_2 \\ U_{L2} \\ \psi_2
\end{array} \right).
\eeq
A linear combination of $\psi_1$ and $\psi_2$ will become the SM doublet.
To decouple the extra components, we need 5 extra fermions:
$\psi_3$, $\psi_4$ and $\psi_5$ are $SU(2)_3$ doublets, and $U_{R1}$ and $U_{R2}$ are
singlets. We also need two extra scalar fields, $F_1\in {\bf 5}$ and $F_2\in {\bf \bar{5}}$ of $SU(5)$. Both are uncharged under $SU(2)_3\times U(1)_3$.
\begin{table}
\begin{center}
\begin{tabular}{|c|ccc|} \hline
& $SU(5)$ & $SU(2)_3$  & $U(1)_3$ \\ \hline
$\Psi_1$ & $\overline{\Yfund}$ & 1 & $Y + 1/2$\\
$\Psi_2$ & $\Yfund$ & 1 & $Y + 1/2  $\\
$\psi_{3}$ & $1$ & $\Yfund$ & $-Y$ \\
$\psi_{4,5}$ & $1 $ & $\Yfund$ & $-Y -1$ \\
$U_{R1,2}$ & $ 1$ & $ 1 $ &   $-Y -1/2$ \\
$u_{R}$ & $ 1$ & $ 1 $ &  $-Y - 1/2 $ \\
$d_{R}$ & $ 1$ & $ 1 $ &  $-Y + 1/2 $ \\ \hline
\end{tabular}
\caption{Fermion fields required to incorporate one generation of SM quarks, and their gauge charge assignments. Here $Y=1/6$ is the SM quark doublet hypercharge. For a generation of leptons, the same set of fields is required, except $d_R\to e_R$, $u_R$ is omitted if the neutrino is Majorana (or $u_R\to \nu_R$ if it is Dirac), and $Y=-1/2$.}
\label{tab:fermions}
\end{center}
\end{table}
Under T-parity,
\beq \begin{split}
\Psi_1 & \leftrightarrow  \Omega^\dagger  \Psi_2 \\
\psi_3 & \rightarrow \omega \psi_3 \\
\psi_4 & \leftrightarrow \omega \psi_5 \\
U_{R1} & \leftrightarrow U_{R2} \\
u_{R} & \rightarrow u_{R} \\
d_{R} & \rightarrow d_{R} \\
F_1 & \leftrightarrow \Omega F_2.
\end{split} \eeq
The Yukawa couplings allowed by gauge symmetries and T-parity are:
\beq
\mathcal{L}_\text{Yuk} = \kappa_1 \left[ \Psi_1 K_1 \psi_3 + \Psi_2 K_2 \psi_3 \right]
+ \kappa_2 \left[ \Psi_1^\dagger K_2 \psi_4^\dagger + \Psi_2^\dagger K_1 \psi_5^\dagger \right]
+ \kappa_3 \left[ \Psi_1 F_1 U_{R1} + \Psi_2 F_2 U_{R2} \right]
+ {\rm~h.c.}.
\label{lightyuk}
\eeq
The invariance under T-parity can be easily shown using  $\Omega^\dagger \Omega = \mathds{1}$ and $\omega^\dagger \omega =\mathds{1}$. This form of the Yukawas, together with the requirement of the correct hypercharges for the SM fields, unambiguously fixes the $U(1)_3$ charges for all fermions. The gauge quantum numbers of the fermions are summarized in Table~\ref{tab:fermions}.

The fundamental scalars get vevs consistent with T-parity:
\beq
\langle F_1 \rangle = \langle F_2 \rangle = (0,0,f_F,0,0)^T\,,
\label{Fvevs}
\eeq
where $f_F\sim$ TeV. These vevs break $Y_1$ and $Y_2$ seperately, but leave $Y_1 + Y_2+Y_3$ unbroken, so that no gauge symmetries not already broken by $S$ and $K$ vevs are broken.

For each SM doublet, our model contains five massive Dirac fermions at the TeV scale\footnote{Note that the T-odd fermion masses are bounded from above by constraints on four-fermion operators~\cite{PEW}, and cannot be much heavier than a TeV.}, three T-odd and the other two T-even. Their masses
are $m_{1-}=\sqrt{2}\kappa_1 f_K$, $m_{2\pm}=\kappa_2 f_K$ and
$m_{3\pm}=\kappa_3 f_F$, where the signs denote the T-parity of each state. There is one massless T-even doublet, $\psi_{SM}=\frac{1}{\sqrt{2}}(\psi_1 - \psi_2)$, which is identified with the SM quark or lepton doublet. In the next subsection, we will explain how the SM Yukawa couplings can be generated in this model.

%%%%%%%%%%%%%%%%%%%%%%%%%%%%
\subsection{The Yukawa couplings} \label{sec:topsector}

We will start with the top Yukawa. Due to the large value of this coupling in the SM, naturalness requires it to be implemented in a way that only breaks the global symmetries of the LHT collectively. It is straightforward to incorporate the top Yukawas of the LHT model in our linear model. For the third generation quarks, we use the set of fields listed in Table~\ref{tab:fermions}. In addition to the terms in~\leqn{lightyuk}, we include the following operators:\footnote{By convention fundamental $SU(5)$ indices are upper, antifundamental are lower. SU(2) indices are raised and lowered with $\epsilon^{ab}$ and $\epsilon_{ab}$ as usual.}
\beq
\mathcal{L}_\text{t} \,=\, \lambda_1 \frac{1}{M}  \left[ \epsilon^{ijk} \epsilon^{xy} \Psi_{1i}
S^{\dagger}_{jx} S^{\dagger}_{ky} +\epsilon_{i'j'} \epsilon_{x'y'z'}  \Psi_2^{x'}
S^{y'i'} S^{z'j'}  \right] u_{R}
+ {\rm~h.c.}
\label{topyuk}
\eeq
where we restrict the summation to $i,j,k \in \{1,2,3\}$, $x,y\in \{4,5\}$ and $i',j'\in \{1,2\}$, $x',y',z'\in \{3,4,5\}$ and $M$ is the mass scale suppressing this dimension-5 operator. Note that eq.~\leqn{topyuk} is T-parity invariant, although this is not immediately manifest; taking the T-parity transformation of the first term yields
\beq \begin{split}
\epsilon^{ijk} \epsilon^{xy} \Psi_{1i} S^{\dagger}_{jx} S^{\dagger}_{ky} & \rightarrow
\epsilon^{ijk} \epsilon^{xy} (\Omega^\dagger \Psi_2)_i (\Omega^\dagger S \Omega^*)_{jx} (\Omega^\dagger S \Omega^*)_{ky} \\
& \qquad = \left[ \epsilon^{ijk45}  \Omega^\dagger_{ix'} \Omega^\dagger_{jy'} \Omega^\dagger_{kz'} \Omega^\dagger_{41} \Omega^\dagger_{52} \right]
\left[ \epsilon^{123xy} \Omega^*_{41} \Omega^*_{52}\Omega^*_{33} \Omega^*_{i'x}  \Omega^*_{j'y} \right] \Psi_2^{x'} S^{y'i'} S^{z'j'} \\
& \qquad = \left[ \epsilon_{x'y'z'}  \det \Omega^\dagger \right]  \left[  \epsilon_{i'j'} \det \Omega^* \right] \Psi_2^{x'} S^{y'i'} S^{z'j'},
\end{split} \eeq
which together with $\det \Omega = 1$ gives exactly the second term
in eq.~\eqref{topyuk}. The expansion to summing over 1 to 5 (and then restricting
again to partial summation as in eq.~\leqn{topyuk}) in this derivation is
possible due to the special structure of $\Omega$. After the $S$ field gets a
vev and the radial modes are integrated out,
eq.~\eqref{topyuk} reduces to the top Yukawa term of the usual nl$\sigma$m 
LHT model (see e.g.~\cite{LHT,PEW,JP}). These Yukawa couplings incorporate
the collecitve symmetry breaking mechanism, which protects the Higgs mass
from large renormalization by top loops.    

We now want to obtain the operators in eq.~\eqref{topyuk} from an
$SU(5)$-invariant, renormalizable Lagrangian.
To restore $SU(5)$ invariance, let us introduce two scalar fields,
\beq
A_1 \in \overline{{\bf10}}\,,~~~A_2 \in {\bf 10}\,,
\eeq
with T-parity action
\beq
A_1 \leftrightarrow \Omega^\dagger A_2 \Omega^*.
\eeq
These fields get vevs
\beq
\langle A_1 \rangle = f_A \,
\left( \begin{array}{ccc} 0 & &  \\ & 0 & \\ & & \varepsilon \end{array} \right),
~~~\langle A_2 \rangle = f_A \,
\left( \begin{array}{ccc} \varepsilon & &  \\ & 0 & \\ & & 0 \end{array} \right),
~~~{\rm where~~} \varepsilon =
\left( \begin{array}{cc} 0 & 1  \\ -1 & 0 \end{array} \right).
\eeq
These vevs do not break T-parity or the gauged $SU(2)$s, but break
the $Y_1$ and $Y_2$ gauged generators. So,  the $A$'s need to be charged under $U(1)_3$ with charges chosen such that the broken linear combinations are orthogonal to the one identified
with hypercharge, $Y_1+Y_2+Y_3$. This requires $Q_3(A_1)= Q_3(A_2)=-1$. In addition to their role in the top sector, the antisymmetric fields also help resolve the problem with the correct value of the Weinberg angle mentioned earlier. For a disussion of this issue, see section~\ref{sec:constraints}.

Eq.~\leqn{topyuk} can now be thought of as the low-energy limit of
the following ($SU(5)$-invariant, but still non-renormalizable) Lagrangian:
\beq \begin{split} \label{topyuksu5}
\mathcal{L}_\text{t} \propto& \, \,
 \left[ \epsilon^{abcde} \Psi_{1a}  S^\dagger_{bx}  S^\dagger_{cy}  (A_{1})_{de} ( A_{1}^*)^{xy} 
+  \epsilon_{abcde} \Psi_2^a S^{bx} S^{cy}   (A_2)^{de} (A^*_2)_{xy} \right] u_{R}
 +~{\rm h.c.},
\end{split} \eeq
where the summations are no longer restricted and run from 1 to 5.

One possible way to obtain a renormalizable model is to introduce four scalar fields, $\eta,
\eta^\prime, \xi$, and $\xi^\prime$. These are uncharged under $SU(2)_3\times
U(1)_3$, and transform under $SU(5)$ as follows:
\beq \label{heavyscalars}
\eta \in \Yfund,~~\eta^\prime \in \overline{\Yfund},~~\xi, \xi^\prime \in
{\rm~Adj}.
\eeq
T-parity acts in by-now familiar way:
\beq
\eta \leftrightarrow \Omega \eta^\prime\,,~~~\xi \leftrightarrow
\Omega^\dagger \xi^\prime \Omega\,.
\eeq
The renormalizable Lagrangian is then given by
\beq\begin{split}
\mathcal{L}_\text{t} & \propto \,
\Psi_{1a} \eta^a u_R + \epsilon^{abcde} \eta^\dagger_a S^\dagger_{bx} \xi^x_{\,\,\,c} (A_{1})_{de} + m_0 (\xi^\dagger)^{\,\,c}_x S^\dagger_{cy} (A^*_{1})^{xy} \\
& \quad +\Psi_{2}^a \eta^\prime_a u_R + \epsilon_{abcde} \eta^{\prime \dagger a} S^{bx} (\xi^{\prime \dagger})_x^{\,\,c} (A_{2})^{de} + m_0  \xi^{\prime x}_{\,\,\,c} S^{cy} ({A}^*_{2})_{xy} +~{\rm h.c.} \\
\end{split} \label{topyukfinal} \eeq
plus mass terms for the scalars. Assuming that the scalars are heavier than
$f$, integrating them out reproduces eq.~\eqref{topyuksu5}.

With the above quantum numbers there is no Yukawa coupling possible for the
leptons and the down quarks, which resembles the top Yukawa in
eq.~\leqn{topyuk}. However, it is possible to write down a dimension-6
operator to generate these Yukawa couplings. For the down quarks, this
operator has the form
\beq \begin{split} \label{downyuk}
\mathcal{L}_\text{d} \sim & \frac{\lambda}{M_d^2} \left( \epsilon_{ijk}
\epsilon_{xy}  \Psi_2^x K_1^{ia} {K}^{j}_{1\,a} S^{ky} +
\epsilon^{i'j'} \epsilon^{x'y'z'}  \Psi_{1i'} K^{\, \, a}_{2x'} K_{2y'a}
{S}^\dagger_{z'j'}  \right) d_{R}\,+\,{\rm h.c.},
\end{split} \eeq
where the summation is restricted to $i,j,k \in \{1,2,3\}$, $x,y\in \{4,5\}$
and $i',j'\in \{1,2\}$, $x',y',z'\in \{3,4,5\}$, and $M_d$ is the  mass scale
at which this operator is generated. The lepton Yukawas are of the same form.
In complete analogy to the top sector, the desired operators can be obtained
from a renormalizable and $SU(5)$ invariant lagrangian by introducing new
heavy states (scalars or fermions) and integrating them out.

%%%%%%%%%%%%%%%%%%%%%%%%%%%%
\subsection{A non $SU(5)$ invariant theory}

One might wonder if the rich structure of the model we built is just due to
the requirement of $SU(5)$ gauge invariance at high energies. If one is willing
to assume that the $SU(5)$ global symmetry accidentally emerges at the 10 TeV
scale, a model with ungauged $SU(5)$ can be considered. Could this dramatically
simplify the particle content needed to reproduce the LHT? A detailed look at
the previous section reveals that only very few states could actually be
omitted in such a non-$SU(5)$ invariant model:
\begin{itemize}
\item We could use incomplete $SU(5)$ representations in~\eqref{su5fermions} and
omit the states $\chi_{1,2}$.
\item We would not need the scalars $F_{1,2}$ to give mass to the $U_{L1,2}$
states.
\item We would not need the scalars $A_{1,2}$, whose role is to make the
coupling~\eqref{topyuk} $SU(5)$ invariant.
\item Fewer massive scalars would be necessary to obtain the top
Yukawas~\eqref{topyuk} from a renormalizable theory.
\end{itemize}
In total one would end up with a slightly smaller particle content, but
overall the model would not simplify significantly.

%%%%%%%%%%%%%%%%%%%%%%%%%%%%%%%%%%%%%%%%%%%%%%%%%%%%%%
%%%%%%%%%%%%%%%%%%%%%%%%%%%%%%%%%%%%%%%%%%%%%%%%%%%%%%
\section{Anomaly Cancellation}
\label{sec:anomalycancel}  \setcounter{equation}{0} %\setcounter{footnote}{0}
%%%%%%%%%%%%%%%%%%%%%%%%%%%%%%%%%%%%%%%%%%%%%%%%%%%%%%
%%%%%%%%%%%%%%%%%%%%%%%%%%%%%%%%%%%%%%%%%%%%%%%%%%%%%%
While the model presented above suffers from gauge anomalies, in this
section we will present a simple extension of the model which is anomaly free.
Furthermore, we will show that T-parity is an anomaly free symmetry of the
quantum theory.

%%%%%%%%%%%%%%%%%%%%%%%%%%%%
\subsection{Gauge anomalies}
First, we examine the gauge anomalies of the model. The chiral fermion content
of a single generation is summarized in Table~\ref{tab:fermions},
where $Y=1/6$ for quarks and $Y=-1/2$ for leptons. Note that the $SU(5)$ group
is vectorlike, while $SU(2)$ representations are real, so all anomalies involving
only these two groups vanish. However, anomalies involving U(1)$_3$ are
{\it not} canceled with this fermion content. The simplest way to achieve
anomaly cancelation is to extend the model in such a way that it contains a
sector which is vectorlike under the full $SU(5)\times SU(2)_3\times U(1)_3$
gauge group, plus a sector which is chiral under $SU(2)_3\times U(1)_3$, but
with charges identical to one generation of the SM fermions. This guarantees
anomaly cancelation as in the SM. Since at low energies the matter content
of our model coincides with the SM, this is in fact possible. In order to
achieve this, we need to introduce mirror partners for all fields that don't
already have SM quantum numbers. In particular for the quark sector we
introduce the mirror partners $Q_1'$, $Q_2'$, $q_4'$, $q_5'$, $U_{R1}'$,
$U_{R2}'$ and \emph{two} fields $q_3'$, $q_3''$. The two $q_3$ partners are
necessary in order to exactly reproduce the chiral SM matter content under
$SU(2)_2 \times U(1)_3$, guaranteeing complete anomaly cancelation. The
total anomaly-free fermion content in the quark sector is summarized in
Table~\ref{tab:completefermions} in the columns (a) and (b).

\begin{table}
\begin{center}
\beq
\begin{array}{|c|ccc} \hline
\text{a)}& SU(5) & SU(2)_3  & U(1)_3 \\
\hline
Q_1 & \bar{\Yfund} & 1            & +2/3 \\
Q_2 & \Yfund          & 1           &  +2/3 \\
q_3 & 1                   &  \Yfund & - 1/6  \\
q_4 & 1                 &  \Yfund &  -7/6  \\
q_5 & 1                 &  \Yfund &  -7/6 \\
U_{R1} & 1              & 1             & -2/3 \\
U_{R2} & 1             & 1             & -2/3 \\
u_{R} & 1              & 1            & -2/3 \\
d_{R} & 1              & 1            & +1/3 \\ \hline
\end{array}
\hspace{-0.1cm}
\begin{array}{|c|ccc} \hline
\text{b)}& SU(5) & SU(2)_3  & U(1)_3 \\
\hline
Q'_1 & \bar{\Yfund} & 1            & -2/3 \\
Q'_2 & \Yfund          & 1           &  -2/3 \\
q'_3, q''_3 & 1                   &  \Yfund & + 1/6  \\
q'_4 & 1                 &  \Yfund &  +7/6  \\
q'_5 & 1                 &  \Yfund &  +7/6 \\
U'_{R1} & 1              & 1             & +2/3 \\
U'_{R2} & 1             & 1             & +2/3 \\
    &   &   & \\
    &   &   & \\ \hline
\end{array}
\hspace{-0.1cm}
\begin{array}{||c|ccc|} \hline
\text{c)}& SU(5) & SU(2)_3  & U(1)_3 \\
\hline
L_1 & \bar{\Yfund} & 1            & 0 \\
L_2 & \Yfund          & 1           &  0 \\
\ell_3 & 1                   &  \Yfund & + 1/2  \\
\ell_4 & 1                 &  \Yfund &  -1/2  \\
\ell_5 & 1                 &  \Yfund &  -1/2 \\
E_{R1} & 1              & 1             & 0 \\
E_{R2} & 1             & 1             &  0 \\
e_{R} & 1              & 1            & +1 \\
(\nu_{R} & 1              & 1            & 0\hspace{2mm}) \\ \hline
\end{array} \nonumber
\eeq
\caption{The complete fermion sector (single generation) and the gauge charge
assignments for the anomaly-free version of the model.}
\label{tab:completefermions}
\end{center}
\end{table}

The additional states acquire TeV-scale masses through a Lagrangian of the form
\beq \label{mirrorByuk}
{\cal L} \propto Q_1' K_2^* q_3' + Q_2' K_1^* q_3'' +{ Q_1'}^\dagger  K_1^* {q_4'}^\dagger+ {Q_2'}^\dagger K_2^* {q_5'}^\dagger + Q_1' F_1 U_{R1}' + Q_2' F_2 U_{R2}'\,.
\eeq
Note that this is almost the same as eq.~\eqref{lightyuk}, except that the
presence of the two \emph{different} fields $q_3'$ and $q_3''$ guarantees
that there is no light mode.

For the lepton sector with $Y=-1/2$ in Table~\ref{tab:fermions} we
automatically have a charge assignment that produces the SM chiral matter
content under $SU(2)_3\times U(1)_3$, so no additional mirror fields are
needed. The matter content in the lepton sector is summarized in
Table~\ref{tab:completefermions} (c).

\begin{table}
\begin{center}
\begin{equation*}
\begin{array}{|c|cccc|} \hline
& SU(5) & SU(3)_c &  SU(2)_3  & U(1)_3 \\
\hline
q''_3 & 1   & \Yfund       &  \Yfund & + 1/6  \\
u_{R}& 1 & \bar{\Yfund} &    1    &  -2/3 \\
d_{R}& 1 & \bar{\Yfund} &    1    &  +1/3 \\
\ell_5 & 1     &    1       &  \Yfund &  -1/2 \\
e_{R} & 1     & 1       & 1            & +1 \\
\hline
\end{array}
\end{equation*}
\caption{The chiral matter content for one generation of the anomaly-free
version of the model.}
\label{tab:chiral}
\end{center}
\end{table}

The chiral matter content of one generation of the model is summarized in
Table~\ref{tab:chiral}.
Here  $SU(3)_c$ denotes the color gauge group. As anticipated above, the
quantum numbers of these fermions under $SU(3)_c \times SU(2)_3
\times U(1)_3$ are exactly the same quantum numbers as for the usual SM
fermions under $SU(3)_c \times SU(2)_L \times U(1)_Y$. Hence all gauge and
gravitational anomalies cancel.

The above construction should be viewed as a proof of principle,
showing that it is possible to add a set of spectator fermions to
our model to cancel all gauge and gravitational anomalies, and to give
them large masses in a way consistent with the symmetries.
The particular set of spectators chosen here is rather large, but has the
advantage that the anomalies cancel in exactly the same way as in the SM.
Its disadvantage is that the QCD $\beta$-function will become very large and the theory would rapidly develop a Landau pole. The exact location of the pole depends on the values chosen for the Yukawa couplings and vevs in eqs.~\eqref{mirrorByuk} and~\eqref{lightyuk}. In the supersymmetric version of this model, which we will describe in section~\ref{sec:HEcompletionsSUSY}, this implies that once the Landau pole is hit an appropriate Seiberg duality~\cite{Seiberg:1994pq} has to be performed and the theory will be a cascading gauge theory as in~\cite{Klebanov:2000hb}. It would be interesting to see if a more minimal anomaly-free matter content
can be found.

%%%%%%%%%%%%%%%%%%%%%%%%%%%%
\subsection{T-parity anomalies}

Whenever physical Goldstone bosons appear in a theory, one has to
check whether the global symmetries whose spontaneous breaking
produces the Goldstones are anomalous. The presence of such
anomalies would produce new couplings for the Goldstones, of the
general form
\beq
\frac{1}{f} \pi^a \partial_\mu J^{a\mu}\,.
\eeq
If the global current $J^{\mu a}$ is anomalous with respect to a
gauge symmetry, then
\beq
\partial_\mu J^{a\mu} = \frac{A g^2}{16 \pi^2} {\rm Tr} F\tilde{F}\,,
\eeq
where $F$ is the gauge field, and the anomaly coefficient $A$
can be calculated from the triangle diagrams involving fermion
loops. In the low energy effective theory after the fermions are
integrated out, a term involving the light gauge fields and the
Goldstones has to be present, whose variation reproduces the
anomalies of the global current. This is the Wess-Zumino-Witten (WZW) term \cite{WZW},
whose coefficient can be found by matching to the triangle diagrams
in the high energy theory. This WZW term may break 
discrete symmetries of the Goldstone sector. The canonical example is
the $\pi^a\to -\pi^a$ symmetry of the pseudoscalar octet of QCD.
The effect of the $SU(2)_A^2$ $U(1)_{em}$ anomaly in the quark picture will imply the presence of the $\pi_0 F\tilde{F}$ coupling in the effective low-energy theory, which breaks the $\pi \to -\pi$ reflection symmetry. Using similar arguments  Hill and Hill~\cite{Hill2} argued
that T-parity
will also be broken in a similar way in little Higgs models. They
have discussed several examples based both on more complicated
versions of the $SU(3) \times SU(3) \to SU(3)_D$ breaking pattern,
as well as the $SU(5) \to SO(5)$ and other little Higgs-type models,
and have calculated the form of the Wess-Zumino-Witten terms in a variety
of examples. However, whether these T-parity breaking terms are
ultimately present in the low-energy effective theory or not depends
on the UV completion of the theory. If the global symmetries (and
T-parity itself) are not anomalous, then the coefficient of the
Wess-Zumino term vanishes, and T-parity remains a good symmetry at the
quantum level. Therefore, in a complete model with T-parity one has
to show that T-parity is not broken by any of the global anomalies
present in the theory. While in an effective low-energy theory one
may only speculate whether such anomalies are present or not, our UV
completion allows us to address this issue straightforwardly. Since
the $SU(5)$ global symmetry responsible for producing the Goldstones
is also gauged, it has to be anomaly free. Indeed we have
shown above that it is possible to choose the matter content such
that all anomalies involving $SU(5)$ will disappear.  Therefore there
can be no Wess-Zumino-Witten term from $SU(5)$ anomalies present in this theory that would give
rise to T-parity violation.

A final worry might be that the T-parity itself as a discrete symmetry might
be anomalous. However, as we have seen before, T-parity is a combination of
an $SU(5) \times SU(2)_3$ gauge transformation element with a discrete
exchange symmetry. We have seen that the gauge transformations are anomaly
free, but what about the exchange symmetry (which is a symmetry similar to
charge conjugation)? Could that possibly be anomalous? The answer is clearly
negative. The exchange symmetry in the path integral language merely
corresponds to a relabeling of the integration variables. The integration
measure is invariant under this relabeling. So, if the Lagrangian is invariant
under the exchange symmetry, then the whole path integral is invariant.
Therefore we do not expect T-parity violating anomalous terms to show up
anywhere in the model.

%%%%%%%%%%%%%%%%%%%%%%%%%%%%%%%%%%%%%%%%%%%%%%%%%%%%%%
%%%%%%%%%%%%%%%%%%%%%%%%%%%%%%%%%%%%%%%%%%%%%%%%%%%%%%
\section{Solutions to the Large Hierarchy Problem}
\label{sec:HEcompletions}  \setcounter{equation}{0} %\setcounter{footnote}{0}
%%%%%%%%%%%%%%%%%%%%%%%%%%%%%%%%%%%%%%%%%%%%%%%%%%%%%%
%%%%%%%%%%%%%%%%%%%%%%%%%%%%%%%%%%%%%%%%%%%%%%%%%%%%%%

We constructed a weakly coupled, four-dimensional UV completion of
the LHT model, with T-parity exact at the quantum level. However, the model
assumes a large hierarchy between the scale of scalar vevs (1 or 10 TeV),
and the Planck scale. This hierarchy needs to be stabilized. In this section,
we will explore two possible ways this can be achieved: by embedding the
model into a supersymmetric theory above 10 TeV, and by promoting it to a
warped-space five-dimensional model with the Planck scale at the infrared
(IR) boundary of order 10 TeV.

%%%%%%%%%%%%%%%%%%%%%%%%%%%%
\subsection{A supersymmetric version}
\label{sec:HEcompletionsSUSY}  

It is straightforward to supersymmetrize our model by promoting all fields to
superfields, and assuming that the components that do not appear in our model 
receive soft masses at the 10 TeV scale. In addition, one needs to introduce 
a superfield $\bar{S}$, which has the same quantum numbers as $S^\dagger$. 
This fields gets interchanged with $S$ under T-parity in the familiar way 
$S \leftrightarrow \Omega \bar{S} \Omega^T$. It ensures that it is possible 
to write down a superpotential that allows for the vev in eq.~\eqref{svev1}
and generates the Yukawa couplings~\eqref{topyukfinal}.
We assume the superpotential of the form  
\beq
W = W_\Phi(\Phi_1,\Phi_2)\,+\, W_{\rm Yuk}(S,\bar{S}, K_1, K_2,
\ldots)\,,
\eeq
where $W_\Phi$ generates $SU(5)$ breaking vevs as in 
eq.~\leqn{phivev} without breaking SUSY, and $W_{\rm Yuk}$ includes the Yukawa 
couplings of our model. This superpotential allows for the adjoint 
vevs in Eq.~\leqn{phivev}, with $\langle \sigma \rangle = 0$. At the same 
time, since the Yukawa couplings do not
contain the $\Phi$ fields, it does {\it not} lead to direct couplings 
between $\Phi$ and the other fields in the F-term scalar potential. As a 
result, the global SU(5) symmetry below the scale $f_\Phi\sim 10$ TeV is 
preserved at this level. Note that this structure of the F-term potential is 
technically natural, due to the standard non-renormalization theorems of SUSY. 

The scalar potential also receives a D-term contribution. 
Since both $\Phi$ and the other scalar fields, including $S$ and $\bar{S}$, are charged
under SU(5), the D-term potential will in general couple them, violating 
the global SU(5). This can give a large contribution to the Higgs mass,
potentially of order $g_5f_\Phi$. However, it can be shown that this effect 
is suppressed 
in the limit when the soft masses for the adjoint fields are small compared
to $f_\Phi$, and the Higgs mass can remain at the weak scale without 
fine-tuning.

The argument is based on the following  observation~\cite{Kawamura:1994ys,Adamnotes}: In the limit of unbroken
SUSY, the effective theory below the scale $f_\Phi$ is a supersymmetric theory 
with reduced gauge 
symmetry. This SUSY theory does {\it not} contain any D-terms for $S$ or $\bar{S}$ 
corresponding to the broken generators, and does not contain any 
$\Phi$ fields as they are either eaten or get masses at the scale $f_\Phi$.
So, in this limit we are only left with D-terms for $S$ and $\bar{S}$ corresponding to the 
unbroken subgroup. These terms do not generate a tree-level $S$ or $\bar{S}$ mass, and 
moreover they break the SU(5) in exactly the same pattern as the unbroken 
gauge symmetries themselves. In particular, the Higgs (contained in $S$ and $\bar{S}$) would 
still remain a Goldstone if only one of the two $SU(2)$ subgroups was gauged.
Thus, in the unbroken-SUSY limit, the D-terms do not spoil the symmetries 
responsible for keeping the Higgs light.

Let us see explicitly how this works. Since for the protection of the higgs mass only the interactions between $S,\bar{S}$ and $\Phi_{1,2}$ are relevant, we will only focus on these fields on the following discussion. Above $f_\Phi$, the D-term 
potential has the form 
\beq \begin{split} \label{Dterms}
V_D &= \frac{g_5^2}{2} \sum_a (D_\Phi^a  + D_S^a + ... )^2, \\
 \text{with } & \quad D^a_\Phi = \sum _i {\rm Tr}\,\Phi_i^\dagger [ T^a, \Phi_i]\,,~~~~
D^a_S \,=\, 2 \, {\rm Tr}\,S^\dagger T^a S - 2 {\rm Tr} \bar{S}^\dagger T^{aT} \bar{S}\,.
\end{split} \eeq 
After the $\Phi$'s get vevs, this potential includes $SU(5)$ symmetry breaking terms
for $S$ and $\bar{S}$. However, to obtain the correct low-energy potential, we 
have to carefully integrate out the heavy ``radial'' modes of the $\Phi$
fields. The important radial modes are $R^{\hat{a}}$ along the generators 
$T^{\hat{a}}$ broken by $\langle \Phi_{1,2} \rangle$. These modes are the 
real parts of the superfield containing the Goldstones, and as such they 
must be F-flat directions.\footnote{Non-linearly realized Goldstones are 
completely F-flat. If realized linearly, however, one will encounter quartic 
and higher interactions in the F-term potential.} But since the Goldstones 
are eaten by 
the broken gauge bosons, the $R^{\hat{a}}$ fields will get masses from the 
D-terms, which must be precisely equal to the gauge boson masses in order to 
preserve SUSY. Furthermore, they are the only radial modes that receive a 
mass from the D-terms. The scalar potential has the form
\beq \label{SUSYpot}
V_\text{SUSY} = F^* F +   \frac{g^2}{2} D^a D^a = 
\frac{1}{2} \sum_{\hat{a}} ( M_{\hat{a}} R^{\hat{a}} + ... +  g_5 D_S^{\hat{a}})^2 
\, + ...  \, , 
\eeq
where $\hat{a}$ labels the broken generators, $M_{\hat{a}}$ are the gauge 
boson masses and the dots denote terms that do not 
contain either $D_S^{\hat{a}}$ or $R^{\hat{a}}$. The equations of motion 
yield
\beq \label{eomforR_susy}
R^{\hat{a}} = -\frac{g_5 D_S^{\hat{a}}}{M_{\hat{a}}}\,,
\eeq
which exactly cancels the unwanted D-terms for $S$ and $\bar{S}$ corresponding to the 
broken generators. 

In a realistic model, SUSY must be broken. Consider a situation when the
SUSY-breaking soft masses for the $\Phi$ fields are lower than the $SU(5)$ 
breaking scale $f_\Phi$. Assume that the soft breaking are of the form
\beq \label{nonSUSYpot}
V_\text{\noSUSY} = \frac{1}{2} \sum_{\hat{a}}  m_{\hat{a}}^2  {R^{\hat{a}}}^2 + \ldots,
\eeq
with $m_{\hat{a}} \ll f_\Phi$, and dots denote terms not containing 
$R^{\hat{a}}$.
The important feature of these soft terms is that they do 
not contain a linear term in $R^{\hat{a}}$, and thus only affect 
the SUSY cancellation of the D-terms at subleading order in $m_{\hat{a}}/M_{\hat{a}}$.
The equations of motion for $R^{\hat{a}}$ now yield
\beq \label{eomforR}
R^{\hat{a}} = -\frac{g_5 D_S^{\hat{a}} M_{\hat{a}}}{M^2_{\hat{a}}+m^2_{\hat{a}}} +...  \approx -\frac{g_5 D_S^{\hat{a}}}{M_{\hat{a}}} \left( 1 + \frac{m^2_{\hat{a}}}{M^2_{\hat{a}}} + ... \right)  \,.
\eeq
The resulting low-energy potential has the form
\beq
V_\text{eff} \sim \sum_{\hat{a}}  \frac{m^2_{\hat{a}}}{M^2_{\hat{a}}} \left( g_5 D_S^{\hat{a}} \right)^2 + \ldots
\eeq
where the dots denote terms of higher order in $m_{\hat{a}}/f_\Phi$. This 
potential gives a mass to the Goldstones in $S$ and $\bar{S}$ (including the SM Higgs) of 
the order 
\beq
m_h^2 \sim \frac{m^2_{\hat{a}}}{M^2_{\hat{a}}} f_S^2\,.
\label{nonsu5mh}
\eeq
This is phenomenologically acceptable as long as $m_{\hat{a}}/M_{\hat{a}} \lsim 0.1$.
One possibility is that $f_\Phi\sim M_{\hat{a}} \sim 10$ TeV as previously assumed, but the 
soft masses for $\Phi$ are an order of magnitude smaller than the other soft 
masses in the theory, $m_{\hat{a}} \sim 1$ TeV. This small mass hierarchy would be 
radiatively stable. Another possibility is that $m_{\hat{a}} \sim 10$ TeV along 
with the other soft masses, but $f_\Phi\sim 100$ TeV. In this case, 
all quadratic divergences are still cut off at 10 TeV due to SUSY, but
SU(5)-violating logarithmic corrections are enhanced by running between 
10 and 100 TeV scales. This leads to an additional contribution to the Higgs 
mass of order $\sim \frac{g^2}{16 \pi^2} f_S^2 \log \frac{100\text{ TeV}}{10\text{ TeV}}$, which is of the same order as the top contribution.
  
The above discussion is completely general and does not depend on any 
particular representation of the $SU(5)$ breaking fields and their vevs, 
the specific form of the superpotential $W_\Phi$, or the soft breaking 
potential $V_{\noSUSY}$. As an example consistent with our model, 
we can use a T-parity invariant superpotential of the form
\beq
W = \kappa \sigma (\Tr \Phi_1 \Phi_1 + \Tr \Phi_2 \Phi_2 - 60 f_\Phi^2)\,+\, W_{\rm Yuk}(S,\bar{S}, K_1, K_2,\ldots),
\eeq
with $\sigma$ a gauge-singlet chiral superfield, and the soft breaking terms
\beq
V_{\noSUSY} = M_\Phi^2 \left( \Tr \Phi_1^\dagger \Phi_1 + \Tr \Phi_2^\dagger \Phi_2 \right) + M_\sigma^2 |Ê\sigma |^2.
\eeq
This potential has an extended $SU(5)^2$ global symmetry, and thus not all Goldstone bosons are eaten by the heavy gauge field.  However, the uneaten Goldstones will receive a contribution to their mass of order $\frac{f_\Phi}{4 \pi}$ at one loop, which is of order $1-10$ TeV.

%%%%%%%%%%%%%%%%%%%%%%%%%%%%
\subsection{A five-dimensional version}

A popular alternative to supersymmetry for solving the weak/Planck
hierarchy problem is the warped-space five-dimensional (5D) setup
pioneered by Randall and Sundrum~\cite{RS}. It is straightforward to
embed our model into such a setup.\footnote{A 5D version of the
original Littlest Higgs model was given in~\cite{ThalerYavin}.}

\begin{figure}
  \centering
  \includegraphics[width=8cm,keepaspectratio=true]{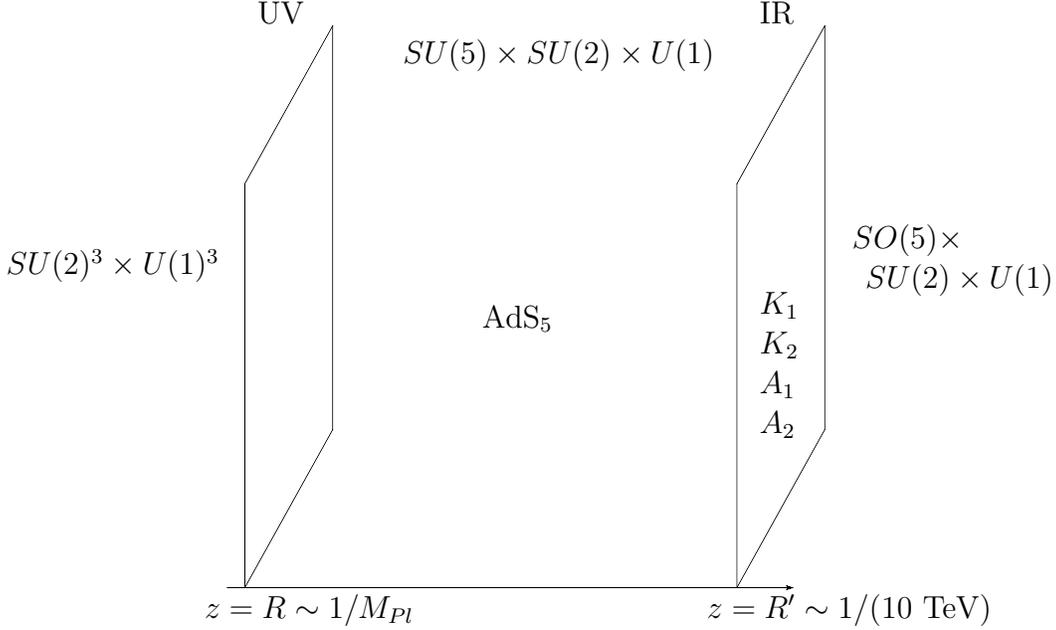}
  \put(-130,100){AdS$_5$}
  \put(-160,200){$SU(5)\times SU(2) \times U(1)$}
  \put(-310,120){$SU(2)^3\times U(1)^3$}
  \put(10,130){$SO(5)\times$}
  \put(15,115){$SU(2)\times U(1)$}
  \put(-25,105){$K_1$}
  \put(-25,90){$K_2$}
  \put(-25,75){$A_1$}
  \put(-25,60){$A_2$}
  \put(-215,215){UV}
  \put(-25,215){IR}
  \put(-235,-10){$z=R\sim 1/M_{Pl}$}
  \put(-45,-10){$z=R'\sim 1/(10 \text{ TeV})$}
  \caption{\label{fig:branes} Geometric setup, gauge symmetries and matter
content of the five-dimensional model.}
\end{figure}

The five-dimensional version of the model is illustrated in
Fig.~\ref{fig:branes}. We assume that the extra dimension has a warped AdS$_5$
gravitational background given by the metric
\beq
ds^2= \left(
\frac{R}{z} \right)^2 \left( \eta_{\mu \nu} dx^\mu dx^\nu - dz^2
\right),
\eeq
The extra dimension is an interval bounded at $z=R$ by the ``ultraviolet'' (UV)
boundary (or brane), and at $z=R'$ by the ``infrared'' (IR) brane.
The AdS curvature $R$ is assumed to be $1/R \sim \mathcal{O}(M_{Pl})$, while
$1/R'$ is of order a few TeV.

The 5D theory should reproduce at $\sim 1$ TeV the T-odd particle
spectrum necessary for the little Higgs mechanism. The cutoff scale
of the 4D little Higgs theory is usually at around 10 TeV. In the 5D
theory this will be identified with the scale $m_{KK}$ where the
additional KK resonances appear, thus UV completing the theory
above 10 TeV. The cutoff scale of the 5D theory can be
estimated via NDA to be of the order $\Lambda_{5D} \sim 24
\pi^3/(g^2 R' \log R'/R) $, while the scale $f$ is given by $f= 2/(g
R' \sqrt{\log R'/R})$. In our case we want $f\sim 1$ TeV, then the
cutoff scale is of order 100 TeV, while the KK mass scale is
$m_{KK}\sim 2/R' \sim 10$ TeV.

The best handle for finding the right setup is to use the dictionary
of the AdS/CFT correspondence. From that point of view we would be
looking for the dual of a CFT with an $SU(5)$ global symmetry, where
the $SU(2)^2 \times U(1)^2$ subgroup is gauged. As we discussed in
this paper, this symmetry needs to be extended to $SU(5)\times
SU(2)_3 \times U(1)_3$, with $[SU(2) \times U(1)]^3$ gauged, in
order to incorporate T-parity in the (chiral) fermion sector. So,
the 5D setup we start with is an $SU(5)\times SU(2)_3 \times U(1)_3$
bulk gauge group. The action of T-parity on the gauge bosons is
again given by eq.~\eqref{Tparity}. We assume that the gauge
symmetry is broken by boundary conditions (BC's) for the gauge
fields, as in~\cite{Higgsless}: on the UV brane, 
\beq 
SU(5) \times
SU(2) \times U(1) \to [SU(2) \times U(1)]^3~~~{\rm (UV)}\,, \\
\label{uv} 
\eeq
while on the IR brane
\beq
SU(5)  \times SU(2) \times
U(1) \rightarrow SO(5) \times SU(2) \times U(1)~~~{\rm (IR)}.
\label{ir}
\eeq 
In the language of the 4D model, this is equivalent
to placing the $\Phi_{1,2}$ fields on the UV brane and the $S$ field
on the IR brane, and integrating out the radial models of these
fields after they get vevs. (Note that this geometric separation of
$\Phi$ and $S$ automatically guarantees the absence of the direct
potential couplings between them, as needed in our model.) These
BC's result in an unbroken $[SU(2)\times U(1)]^2$ gauge group at low
energies and leave T-parity unbroken. The gauge fields in $[SU(2)
\times U(1)]^3$ which are only broken by BC's on the IR brane will
get a mass of order $f \sim 1$ TeV. These fields correspond to the
T-odd gauge bosons of the LHT model. As discussed above, the full
Kaluza-Klein (KK) tower starts at the somewhat higher scale $m_{KK}
\sim $ 10 TeV.

To reduce the group further (down to just the SM) we will assume
that the scalars $K_1$, $K_2$ live on the IR brane, getting vevs of
order $m_{KK}\sim 10$ TeV. Furthermore, to incorporate fermion masses in an
$SU(5)$ invariant way, we also add the scalars $A_1$, $A_2$ on the
IR brane,with vevs of order $m_{KK}$. (We will not need to
introduce the scalars $F_{1,2}$ to give masses to $U_{L1,2}$.) Note that 
$m_{KK}\sim 10$ TeV is the natural scale for the vevs on the IR brane. It is 
an order of magnitude 
larger than the vevs for these fields in the 4D version of the model. However, 
these larger vevs do not lead to larger masses for the corresponding massless
gauge bosons: in fact, their contribution to the masses is at most of order 
$g f \sim 1$ TeV. This can be seen by observing that the limit of very large 
vevs is equivalent to breaking gauge symmetries by BC's on the IR brane, which 
produce masses of order $gf$. 

The $A_5$ components of the gauge fields corresponding to the broken
$SU(5)/SO(5)$ generators develop zero modes. These modes, which are
scalars from the 4D point of view, include the weak doublet
identified with the SM Higgs. The Higgs mass is protected by the
collective symmetry breaking mechanism. To see this, consider a
variation of the symmetry breaking pattern in
eqs.~\leqn{uv},~\leqn{ir}, with $SU(5)$ broken down to a {\it
single} $SU(2)\times U(1)$ subgroup on the UV brane. This theory
possesses an $SU(3)$ global symmetry, broken down to $SU(2)$ by the BC's
on the IR brane. The $A_5$ components identified with the Higgs are
the Goldstone bosons of this global symmetry breaking, and as such
are exactly massless. Thus, the Higgs can only get a mass if {\it
both}  $SU(2)\times U(1)$ factors in $SU(5)$ are unbroken at the UV
brane. That is, zero modes for at least two different gauge fields
must enter into any diagram contributing to the Higgs mass. Just as
in the 4D LHT, this implies cancelation of the quadratic divergence
in the Higgs mass between the SM gauge bosons and their T-odd
counterparts at scale $f$. The remaining logarithmic divergence is
canceled by the KK states at the scale of order $1/R^\prime\sim 10$
TeV, and a finite Higgs mass is generated, as guaranteed by non-locality
and 5D gauge invariance. Note that there may be additional light
states among the $A_5$ modes due to the large vevs of $K_{1,2},
A_{1,2}$ on the IR brane. However, those would not be protected by
the collective breaking mechanism, but only by the 5D non-locality,
so their masses would be of the order of $m_{KK}/4\pi \sim 1$ TeV,
rather than the 100 GeV range for the doubly protected physical
Higgs.

It is useful to compare this structure to that of the ``minimal''
holographic composite Higgs model of Agashe, Contino and
Pomarol~\cite{ACP}. In that model, {\it all} divergences in the
Higgs mass are canceled at the same scale, the KK scale
$1/R^\prime$. Precision electroweak (PEW) constraints push this
scale up to at least 3 TeV, and some amount of fine-tuning is needed
to obtain consistent EWSB. In contrast, in our theory, the quadratic
divergence is canceled at the 1 TeV scale by the Little Higgs
mechanism, without any tension with PEW constraints thanks to T
parity. This allows us to push the KK scale to 10 TeV without
fine-tuning. At this scale, the KK states themselves are completely
safe from PEW constraints. Thus, the tension between fine-tuning and
PEW constraints is eliminated. Of course, the price to pay is a
larger symmetry group and matter content.

In principle, the fermion content  of the five-dimensional model
could be simplified compared to the 4D $SU(5)$-invariant model, if
one were to take advantage of the symmetry breaking BC's and simply
project out some of the unwanted zero modes for the fermions (such
as, for example, $U_i$ and $\chi_i$ components of the $\Psi_i$
fields) instead of introducing new states for them to marry.
However, one needs to be careful with this, if T-parity is to be
maintained as an exact symmetry. 5D theories are automatically
anomaly free in the sense that every bulk fermion is actually a 4D
Dirac fermion, and so the theory is always vectorlike. However, once
orbifold projections are introduced, {\it localized} anomalies can
be generated on the boundaries, which would be locally canceled by
an anomaly flow corresponding to the bulk Chern-Simons (CS)
term~\cite{ArkaniHamed:2001is}. These bulk CS terms would contain
the $A_5$ field and thus could violate T-parity similarly to the WZW
operators in the 4D case. In order to avoid such terms, we need to
make sure that there are no localized anomalies in our theory. The
most obvious way of achieving this is by putting a separate bulk
fermion field for every field in Table~\ref{tab:completefermions},
with a zero mode forming a complete $SU(5)$ representation. This
would imply that we pick a $(+,+)$ boundary condition for all the
left handed components, and a $(-,-)$ BC for all the right handed
components. This choice ensures that all localized anomalies cancel
in the same way as in the 4D theory (see
section~\ref{sec:anomalycancel}), and there would be no bulk CS term
appearing. The terms corresponding to the Lagrangian in
eqs.~\eqref{lightyuk} and~\eqref{mirrorByuk} can then be mimicked by
brane localized Yukawa terms involving the $K_1,K_2$ fields on the
IR brane,  and via UV brane localized mass terms of the form $U_{L1}
U_{R1}+ U_{L2} U_{R2}$ (remember that on the UV brane $SU(5)$ is
broken and so these mass terms are not violating gauge invariance,
so we do not need to introduce $F_{1,2}$). If we were to try to
simplify the spectrum by using $(-,+)$ type boundary conditions for
some of the fermions (and introducing fewer bulk fields), we would
end up with a consistent theory, but with a bulk CS-term breaking
T-parity.

In order to obtain Yukawa couplings, we need to make sure that the
zero modes for the right-handed quarks also partly live in the
right-handed component of $U_{L1,2}$. This can be achieved via the
IR brane localized scalars corresponding to $\eta$, $\eta'$, $\xi$, $\xi'$ in eq.~\eqref{heavyscalars}.
A Lagrangian corresponding
to eq.~\eqref{topyukfinal} can be also added to the IR brane, except for
adding mass terms along the pattern of the $\langle S \rangle$
instead of the complete $S$ field (which is allowed due to the symmetry
breaking BC's). The effect of those boundary terms will be to
partially rotate the $u_{R}$ zero mode into $Q_1$, and thus
generate our effective Yukawa coupling. Note, that since all global
$SU(3)_{1,2}$ violating effects are non-local (as they need to involve
both branes), the radiatively generated Higgs potential will be
completely finite. We leave the detailed study of the
EWSB and the phenomenology of the holographic T-parity models to
future investigations.

%%%%%%%%%%%%%%%%%%%%%%%%%%%%%%%%%%%%%%%%%%%%%%%%%%%%%%
%%%%%%%%%%%%%%%%%%%%%%%%%%%%%%%%%%%%%%%%%%%%%%%%%%%%%%
\section{Constraints from the Weinberg Angle, Precision Electroweak Fits,
and Dark Matter}
\label{sec:constraints}  \setcounter{equation}{0} %\setcounter{footnote}{0}
%%%%%%%%%%%%%%%%%%%%%%%%%%%%%%%%%%%%%%%%%%%%%%%%%%%%%%
%%%%%%%%%%%%%%%%%%%%%%%%%%%%%%%%%%%%%%%%%%%%%%%%%%%%%%
The model constructed in sections~\ref{sec:bosons} and~\ref{sec:fermions}
correctly
reproduces the particle content of the SM at low energies. At the TeV scale,
the model reproduces the particle content and couplings of the LHT. This sector
eliminates the little hierarchy problem, and is consistent with precision
electroweak fits as long as $f_S\geq 500$ GeV, and the T-odd partners
of the SM fermion doublets are not too far above the TeV scale~\cite{PEW}.
In addition, our model contains a number of states at the TeV scale that were
{\it not} present in the LHT. These states can produce additional
contributions to precision electroweak observables. While a detailed
analysis of the resulting constraints is outside the scope of this paper,
we would like to briefly discuss the most salient constraint and show that
it can be satisfied.

Most TeV-scale non-LHT states in our model are vectorlike fermions, and their
contributions to PEW observables are small. The dominant
new contribution is from the massive T-even gauge bosons. As discussed in
section~\ref{sec:bosons}, these states can be significantly heavier than the
T-odd gauge bosons, if the gauge couplings of the $SU(2)_3\times U(1)_3$
gauge groups are stronger than that of the $SU(5)$ group. Since the SM Higgs
does not couple to the $SU(2)_3\times U(1)_3$ gauge bosons, the little
hierarchy problem is still solved in this limit, provided that the T-odd
gauge bosons remain sufficiently light. However, as mentioned at the end of
section~\ref{sec:bosons}, the potential problem with this limit is the
Weinberg angle prediction: the SM coupling are related to the $SU(5)\times
SU(2)_3  \times U(1)_3$ gauge couplings via
\beq
\frac{1}{g^2} = \frac{2}{g_5^2}Ê+ \frac{1}{g_3^2} ~~~Ê\text{ and }~~~Ê\frac{1}{g^{'2}} = \frac{6}{5 g_5^2}Ê+ \frac{1}{g_3^{'2}}\,,
\label{gs}
\eeq
so that $\sin^2\theta=5/8$ in the limit $g'_3, g_3\gg g_5$. Is it possible to
satisfy precision electroweak constraints and at the same time reproduce the
experimental value of the Weinberg angle, $\sin^2 \theta_\text{exp} \approx
0.2315$?

The spectrum of the TeV-scale gauge bosons has been discussed in
section~\ref{sec:bosons}, see eqs.~\leqn{su2masses} and~\leqn{u1masses}.
However, these equations did not take into account the effect of the
additional breaking of the $U(1)$ gauge bosons by the vevs of $A_{1,2}$ and
$F_{1,2}$. Including these vevs, the $U(1)$ gauge boson masses are
\beq \label{u1masseswA}
m_{B_\text{even}}^2  =  \tfrac{g'^2_5+2g'^2_3}{4} (f_K^2+16 f_A^2) ~~~
\text{ and }~~~m_{B_\text{odd}}^2 = \tfrac{{g'}^2_5}{100}(10 f_S^2 +f_K^2+
16 f_A^2+32 f_F^2)\,,
\eeq
(where $g'_5=\sqrt{5/3}\, g_5$),while the $SU(2)$ gauge boson masses are still given by eq.~\leqn{su2masses}.
It is convenient to rewrite the gauge boson spectrum and the Weinberg angle
in terms of dimensionless ratios:
\beq \begin{split} \label{relations}
\sin^2 \theta & = \left[ 1+ \frac{1}{5} \cdot \frac{6+5/r^\prime}{2+1/r}
\right]^{-1} \\
\frac{m_{W_\text{even}}^2}{m_{W_\text{odd}}^2} & = \frac{ 1+2r}{1+ 2r_S} \\
\frac{m_{B_\text{odd}}^2}{m_{W_\text{odd}}^2} & =
\frac{1+10r_S+16r_A+32r_f}{60(1+2r_S)}\\
\frac{m_{B_\text{even}}^2}{m_{W_\text{odd}}^2} & =  \left[
\frac{5}{3} + 2 r^\prime \right]  \frac{1+16r_A}{1+2r_S},
\end{split} \eeq
where the ratios are defined as
\beq
r=g_3^2/g_5^2,~r^\prime=g_3^{\prime 2}/g_5^2,~r_S=f_S^2/f_K^2,
~r_A=f_A^2/f_K^2,~r_F=f_F^2/f_K^2.
\eeq

Tree-level shifts in precision electroweak observables can be computed in terms of the T-even gauge boson masses and the coupling constant ratios, $r$ and $r^\prime$. For example,
taking the $Z$ mass, the Fermi constant $G_F$ and the fine structure constant $\alpha$ as inputs, the shift in the $W$ boson mass with respect to
the reference value is given by
\beq
\Delta m_W \equiv m_W- c_w^{\rm ref} m_Z = \frac{m_W}{4}\,
\frac{\pi \alpha}{c_w^2-s_w^2}\,\left( \,\frac{1}{r}
\frac{v^2}{m^2_{W_\text{even}}}\,+\,\frac{5}{3}\frac{1}{r^{\prime}}
\frac{v^2}{m^2_{B_\text{even}}}\right)\,,
\label{mw}
\eeq
where $c_w^{\rm ref}$ is the reference value of the cosine of the Weinberg 
angle, and $v \approx 246$ GeV is the Higgs vev. The
structure of corrections to {\it all} observables is the same as in
eq.~\leqn{mw}: the contributions of the heavy $SU(2)$ states are proportional to
$r^{-1}m_{W_\text{even}}^{-2}$, while those due to the heavy $U(1)$ states are
proportional to $r^{\prime -1}m_{B_\text{even}}^{-2}$. This is because both
the light-heavy gauge boson mixing, and the couplings of the heavy gauge bosons
to light fermions, are inversely proportional to $\sqrt{r}$ or $\sqrt{r^\prime}$.

This structure can be exploited to find the region of parameter space where the corrections are suppressed without fine-tuning. To avoid
large corrections to the Higgs mass from the $SU(2)$ sector, the $W_\text{odd}$ gauge bosons should be light, preferably around 1 TeV or below. At the same time, the $W_\text{even}$ can be much heavier, if the parameter $r$ is large. In this regime, the contribution to precision electroweak observables from the $SU(2)$ sector is suppressed both by the $W_\text{even}$  mass and by its small mixing and couplings to the SM fermions, as noted above. The PEW constraint on the mass of an extra $SU(2)$ boson with SM-strength couplings (such as the KK gauge bosons in models with extra dimensions) is typically around 3 TeV. Using this value and assuming
$m_{W_\text{odd}}=1$ TeV and $f_S=f_K$, we estimate that the $SU(2)$ contributions in our model are sufficiently suppressed if
$r\gsim 2$. The $r$ parameter is limited from above by the requirement that the $SU(2)_3$ not be strongly coupled:
\beq
\frac{g_3^2}{4\pi} \lsim 0.3~~~~\Leftrightarrow r \lsim 5\,.
\eeq
There is a wide rage of values where the model is perturbative and consistent with data.

Once $r$ is fixed, the requirement of getting the correct Weinberg angle fixes $r^\prime$; the range $2<r<5$ corresponds to $0.14\lsim r^\prime \lsim 0.16$, so that the $U(1)$ mixing angle is essentially fixed.
Thus, the $B_\text{even}$ boson {\it cannot} be decoupled by assuming large $\gp3$. Moreover, the couplings of the heavy $U(1)$ gauge boson to the SM fermions are actually enhanced compared to the SM hypercharge coupling. However, its mass is essentially a free parameter, and it can be heavy provided that $f_A\gg f_S, f_K$. For example, assuming again $m_{W_\text{odd}}=1$ TeV and $f_S=f_K$,
the value of $f_A=3f_S$ gives $m_{B_\text{even}}\approx 10$ TeV,
which should be completely safe for precision electroweak fits even with the enhanced coupling. At the same time, for the same parameters and $f_F=f_S$, the T-odd $U(1)$ boson $B_\text{odd}$
has a mass just above 1 TeV, so that the Higgs mass divergence is still canceled at 1 TeV and there is no fine-tuning. Thus, we estimate that in the region
\beq
2\lsim r\lsim 5\,,~~~r^\prime\approx 0.15\,,~~~r_A\gsim 10\,,
\eeq
and all other dimensionless ratios of order one, our model should be consistent with precision electroweak data without fine-tuning in the Higgs mass.

An interesting phenomenological feature of the spectrum needed to satisfy the
constraints is that the $B_\text{odd}$ boson is not necessarily the lightest T-odd
particle (LTP), in contrast to the situation typical in the original LHT
model. Cosmological considerations require that the LTP not be strongly
interacting or electrically charged. In our model, the T-odd partner of the
SM neutrino can also play the role of the LTP. The T-odd neutrino LTP has not
been considered in the previous studies of Little Higgs dark matter, which
focused on the $B_\text{odd}$ as the dark matter candidate. Our model provides
a motivation to analyze this alternative possibility.

In addition to the gauge bosons, several new scalar states appear at the TeV
scale in our model. These include
pseudo-Goldstone bosons which receive a mass at the one-loop order, as well
as the radial excitations of the fields $S$ and $K_{1,2}$. Several of these
states are triplets with respect to the SM weak SU(2). If allowed by T-parity
and hypercharge conservation, gauge interactions will generate terms of the
form $h^\dagger \phi_i h$, where $\phi_i$ are the triplets, in the one-loop
Coleman-Weinberg potential. Such terms do indeed arise for some of the
triplets in our model. Those triplets are forced to acquire vevs, which can
give large corrections to precision electroweak observables. For example,
this effect played an important role in constraining the original littlest
Higgs model without T-parity~\cite{pew_noT}. In our model, the triplet vevs
are not directly related to the magnitude of the Higgs quartic coupling, as
was the case in the LH without T-parity. We expect that it should be possible
to find phenomenologically consistent regions of parameter space where the
triplet vevs are small.

%%%%%%%%%%%%%%%%%%%%%%%%%%%%%%%%%%%%%%%%%%%%%%%%%%%%%%
%%%%%%%%%%%%%%%%%%%%%%%%%%%%%%%%%%%%%%%%%%%%%%%%%%%%%%
\section{Little Higgs Mechanism in the Linear Sigma Model}
\label{sec:LHMinlsm}  \setcounter{equation}{0} %\setcounter{footnote}{0}
%%%%%%%%%%%%%%%%%%%%%%%%%%%%%%%%%%%%%%%%%%%%%%%%%%%%%%
%%%%%%%%%%%%%%%%%%%%%%%%%%%%%%%%%%%%%%%%%%%%%%%%%%%%%%

A key feature of little Higgs models is the protection of the SM Higgs
mass from quadratic divergence at the one-loop level through collective
symmetry breaking. We argued in sections~\ref{sec:bosons}
and~\ref{sec:fermions} that, since our model below the 10 TeV scale
reproduces the nl$\sigma$m LHT, the same cancelations will occur. While
our model has extra states at the TeV scale, the symmetric scalar
field $S$, which contains the SM Higgs, has no direct couplings to those
states. (It is uncharged under the extra gauge group $SU(2)_3\times U(1)_3$
and has no Yukawa couplings other than the top Yukawa already present in the
LHT.) Thus, no new one-loop quadratic divergences arise. This argument
ensures that in our model the little hierarchy problem is resolved in
exactly the same manner as in the LHT. Nevertheless, it is interesting and
instructive to see explicitly how the little Higgs cancelations occur
in our weakly-coupled, UV-complete model. We will do so in this section.

%%%%%%%%%%%%%%%%%%%%%%%%%%%%
%\subsection{The gauge loop}

\begin{figure}
 \centering
  \includegraphics[width=8cm,keepaspectratio=true]{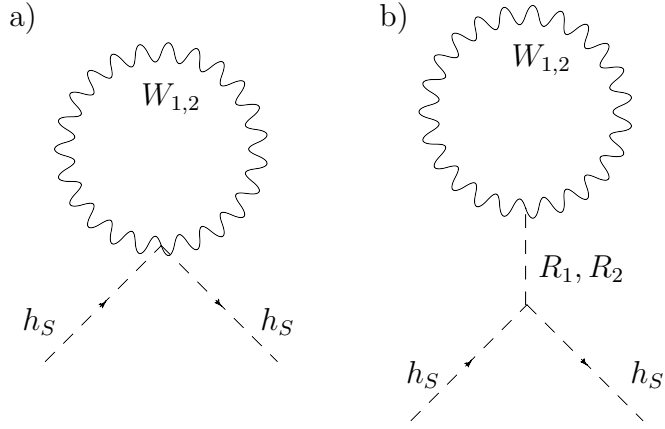}
  % for 1st diagram
  \put(-240,150){a)}
  \put(-145,35){$h_S$}
  \put(-235,35){$h_S$}
  \put(-190,120){$W_{1,2}$}
  % for 2nd diagram
  \put(-100,150){b)}
  \put(-5,15){$h_S$}
  \put(-90,15){$h_S$}
  \put(-50,135){$W_{1,2}$}
  \put(-40,55){$R_1, R_2$}
  \caption{The Feynman diagrams contributing to the effective gauge couplings of the Higgs boson at low energies.}
\label{fig:gaugeloops}
\end{figure}

First, let is consider the renormalization of $h_S$ mass by gauge boson loops.
We will focus on the $SU(2)$ gauge bosons; the analysis for the $U(1)$ bosons
is essentially identical. In our model, the Higgs coupling to the
gauge bosons includes the terms
\beq
\mathcal{L} \supset \tfrac{1}{8} h_S^\dagger h_S \left( g_1^2 W_1^1
+ g_2^2 W_2^2 \right),
\label{bad}
\eeq
where $g_i$ denotes the gauge coupling to the $SU(2)_i$ subgroup of $SU(5)$ (which are the same in our model, but potentially different in the original Littlest Higgs). These terms arise from the covariant derivative in eq.~\leqn{lin} and are required by gauge invariance. These couplings produce a quadratic divergence in the
Higgs mass via the ``bow-tie'' diagrams in Fig.~\ref{fig:gaugeloops} (a).
Recall that in the Littlest Higgs model, the structure of the four-point Higgs-gauge
boson coupling is different~\cite{BPP}:
\beq
\mathcal{L}_\text{LHT} \supset \tfrac{1}{4} g_1 g_2 W_1 W_2 (h^\dagger h),
\label{good}
\eeq
which does not lead to a quadratic divergence at one loop. Since our model
must reduce to the LHT below the 10 TeV scale, there seems to be a
contradiction.

This issue is resolved when the full set of diagrams contributing to the
Higgs mass at one-loop in our linearlized model is included. Specifically,
the relevant diagrams are the ones involving two radial (heavy) modes of $S$,
coupling to the Higgs and the gauge bosons. These diagrams are shown in
Fig.~\ref{fig:gaugeloops} (b). Let us assume that a potential for $S$
has the form
\beq  \label{potentialforS}
V = -M^2 \Tr S S^\dagger + \lambda_1 (\Tr S
S^\dagger )^2 + \lambda_2 \Tr S S^\dagger S S^\dagger,
\eeq
where $M^2=2(5 \lambda_1 + \lambda_2) f_S^2$.  This potential produces the
desired pattern of symmetry breaking at scale $f_S$. It leads to the
following pieces in the Lagrangian containing the heavy radial modes
$R_1$ and $R_2$ (amongst others):
\beq \begin{split} \label{hnRcouplings}
\mathcal{L} \supset &  -\tfrac{1}{2} M^2_{R_1} R_1^2   - \tfrac{1}{2} M^2_{R_2} R_2^2 + \tfrac{1}{\sqrt{5} f_S}  \left( \tfrac{3}{2} M_{R_1}^2  R_1 +   2 M_{R_1}^2  R_2\right) h_S^\dagger h_S \\
& \quad    + \frac{f_S}{4 \sqrt{5}} (R_1 - 2 R_2) \left( g_1^2 W_1^2
+ g_2^2 W_2^2 - 2 g_1 g_2 W_1 W_2 \right)\,,
\end{split} \eeq
where the radial modes have masses $M_{R_1}^2 = 32 \lambda_2 f_S^2$
and $M_{R_2}^2 = 32 (5 \lambda_1 +\lambda_2) f_S^2$. Note that the couplings
of the radial modes to $h_S^\dagger h_S$ are proportional to their masses.
The effective Lagrangian below the scale $f_S$ is obtained by integrating out
the radial modes $R_{1,2}$ in eq.~\leqn{hnRcouplings}. The resulting
Lagrangian contains terms that exactly cancel the gauge-Higgs four-point
couplings in eq.~\leqn{bad}. The remaining coupling has the form
\beq
\mathcal{L}_\text{eff} \supset \tfrac{1}{4} g_1 g_2 W_1 W_2 (h_S^\dagger h_S),
\eeq
which exactly matches the non-linear Littlest Higgs Lagrangian and does not
lead to quadratic divergences at one loop. Note that this result is
independent of the couplings $\lambda_{1,2}$, as expected from the
Coleman-Wess-Zumino theorem.

\begin{figure}
 \centering
  \includegraphics[width=14cm,keepaspectratio=true]{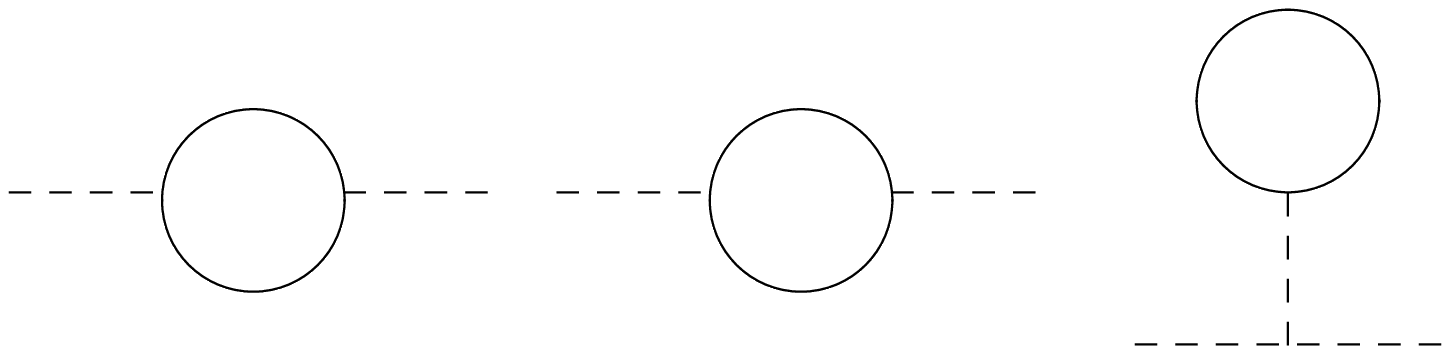}
  % for 1st diagram
  \put(-400,80){a)}
  \put(-350,70){$t_L$}
  \put(-308,12){$t_R$}
  % for 2nd diagram
  \put(-250,80){b)}
  \put(-200,70){$t_L$}
  \put(-158,12){$T_R$}
  % for 3rd diagram
  \put(-90,80){c)}
  \put(-32,62){$T$}
  \put(-40,20){$R_1, R_2$}
  \caption{The Feynman diagrams contributing to the effective top couplings of the Higgs boson at low energies.}
\label{fig:toploops}
\end{figure}

In a completely analogous way, one can show that the diagrams for canceling
the top loop divergence are generated by integrating out $R_1$, $R_2$
properly. These diagrams are shown in Fig.~\ref{fig:toploops}. Especially, we also recover the sum rule from~\cite{PPP} for the Yukawa coupling of the top quark with itself $\lambda_t$ and with its heavy partner $\lambda_T$
\beq
\frac{M_T}{f_S} = \frac{\lambda_t^2+ \lambda_T^2}{\lambda_T},
\eeq
which ensures that the one-loop quadratic divergence due to the top quark cancel.

%%%%%%%%%%%%%%%%%%%%%%%%%%%%%%%%%%%%%%%%%%%%%%%%%%%%%%
%%%%%%%%%%%%%%%%%%%%%%%%%%%%%%%%%%%%%%%%%%%%%%%%%%%%%%
\section{Conclusions and Outlook}
\label{sec:conclusions}  \setcounter{equation}{0} %\setcounter{footnote}{0}
%%%%%%%%%%%%%%%%%%%%%%%%%%%%%%%%%%%%%%%%%%%%%%%%%%%%%%
%%%%%%%%%%%%%%%%%%%%%%%%%%%%%%%%%%%%%%%%%%%%%%%%%%%%%%

In this paper, we constructed a weakly coupled, renormalizable theory which
reproduces the structure of the LHT model below the 10 TeV scale. This
structure includes collective symmetry breaking mechanism to protect the
Higgs mass from one-loop quadratic divergences, resolving the little
hierarchy problem. The model is manifestly free of anomalies, and
T-parity is an exact symmetry of the quantum theory. This leads to an
exactly stable lightest T-odd particle, which can be either the T-odd
hypercharge gauge boson or the partner of the neutrino. This particle can
play the role of dark matter, and provide a missing energy signature at
colliders. In addition, our model contains a few T-even extra states at the
TeV scale, which can however be made sufficiently heavy to avoid conflict
with precision electroweak data, without any fine tuning. Above the 10 TeV
scale, our model can be embedded into either a supersymmetric theory or a
five-dimensional setup with warped geometry, stabilyzing the large hierarchy
between 10 TeV and the Planck scale. A remaining concern regarding the fully anomaly free matter content is that due to the large numbers of states required for anomaly cancelation a Landau pole in the QCD $\beta$-function would develop. It would be very interesting to find a smaller anomaly canceling matter content that can avoid this issue.

In a weakly coupled UV completion of the LHT, a number of issues can be
addressed which could not be analyzed in the original effective theory.
One issue is gauge coupling unification, since in our model renormalization
group evolution of all couplings is calculable all the way up to the
Planck scale. The other one is flavor physics, in particular flavor-changing
neutral currents (FCNCs). There are two sources of FCNCs in the LHT model.
The first one is the effects generated by loops of heavy T-odd quarks and 
leptons, calculable within the effective theory. These effects have been 
considered in~\cite{JayF,Burasz}. The second class are the effects 
generated at or above the cutoff scale of the effective theory. These effects 
should be represented by local operators in the effective theory, with 
coefficients obtained by matching to the UV completion at the cutoff scale.
If the UV completion does not contain any flavor structure, one expects such
operators to appear suppressed by powers of the cutoff scale, with order-one 
coefficients. In the LHT, the cutoff scale is 10 TeV, so several of these 
operators would strongly violate experimental bounds on the FCNCs.  
This indicates that additional flavor structure (e.g. flavor symmetries) 
is a necessary part of the UV completion of the LHT. It would be interesting
to extend out model to obtain realistic flavor physics.

%%%%%%%%%%%%%%%%%%%%%%%%%%%%%%%%%%%%%%%%%%%%%%%%%%%%%%
\section*{Acknowledgments}
%%%%%%%%%%%%%%%%%%%%%%%%%%%%%%%%%%%%%%%%%%%%%%%%%%%%%%

We thank Kaustubh Agashe and Manuel Toharia for discussions leading to this
project. We also thank Hsin-Chia Cheng and Thomas Gregoire for useful 
discussions and comments. 
We are grateful to Adam Falkowski for sharing his unpublished
notes, Ref.~\cite{Adamnotes}, and to David Krohn and Itay Yavin for sending
us an advance copy of their paper~\cite{KY} prior to publication.

Our research is supported by the National Science Foundation under
Grants No. PHY03-55005 at Cornell and PHY05-51164 at the KITP.

\end{document}